# Electrically programmable magnetic coupling in an Ising network exploiting solid-state ionic gating

Chao Yun[1,2,†], Zhongyu Liang[1,†], Aleš Hrabec[3,4,5], Zhentao Liu[3,4], Mantao Huang[6], Leran Wang[1], Yifei Xiao[7], Yikun Fang[7], Wei Li[7], Wenyun Yang[1], Yanglong Hou[2], Jinbo Yang[1], Laura J. Heyderman[3,4,*], Pietro Gambardella[5,*], Zhaochu Luo[1,*]

[1]State Key Laboratory of Artificial Microstructure and Mesoscopic Physics, School of Physics, Peking University, 100871 Beijing, China.
[2]School of Materials Science and Engineering, Peking University, 100871 Beijing, China.
[3]Laboratory for Mesoscopic Systems, Department of Materials, ETH Zurich, 8093 Zurich, Switzerland.
[4]Laboratory for Multiscale Materials Experiments, Paul Scherrer Institut, 5232 Villigen PSI, Switzerland.
[5]Laboratory for Magnetism and Interface Physics, Department of Materials, ETH Zurich, 8093 Zurich, Switzerland.
[6]Department of Materials Science and Engineering, Massachusetts Institute of Technology, Cambridge, MA, USA
[7]Division of functional Materials, Central Iron and Steel Research Institute Group, 100081 Beijing, China.
†These authors contributed equally: Chao Yun, Zhongyu Liang
*Correspondence to: zhaochu.luo@pku.edu.cn (Z.Luo); pietro.gambardella@mat.ethz.ch (P.G.); laura.heyderman@psi.ch (L.J.H.).

**Abstract:**

Two-dimensional arrays of magnetically coupled nanomagnets provide a mesoscopic platform for exploring collective phenomena as well as realizing a broad range of spintronic devices. In particular, the magnetic coupling plays a critical role in determining the nature of the cooperative behaviour and providing new functionalities in nanomagnet-based devices. Here, we create coupled Ising-like nanomagnets in which the coupling between adjacent nanomagnetic regions can be reversibly converted between parallel and antiparallel through solid-state ionic gating. This is achieved with the voltage-control of magnetic anisotropies in a nanosized region where the symmetric exchange interaction favours parallel alignment and the antisymmetric exchange interaction, namely the Dzyaloshinskii-Moriya interaction, favours antiparallel alignment. Applying this concept to a two-dimensional lattice, we demonstrate a voltage-controlled phase transition in artificial spin ices. Furthermore, we achieve an addressable control of the individual couplings and realize an electrically programmable Ising network, which opens up new avenues to design nanomagnet-based logic devices and neuromorphic computers.

**Main text**

The ability to electrically manipulate magnetism is crucial for spintronic applications including spin-based data storage and computation devices. The successful switching of the magnetization in nanomagnets by means of spin transfer torques[1], spin-orbit torques[2-4], voltage-controlled magnetic anisotropy (VCMA)[5-8] and magnetoelectric coupling[9,10] have led to significant steps towards next-generation low-power and high-speed magnetic memories. However, while nanomagnet-based data storage relies on this magnetization switching, nanomagnet logic requires the engineering of the magnetic coupling that aligns the magnetization of adjacent nanomagnets with a specific relative orientation[11-14]. In addition, with the possibility to construct various two-dimensional (2D) coupled nanomagnetic networks, often referred to as artificial spin ices[15,16], the control of lateral coupling is of particular interest for the investigation of collective phenomena such as magnetic frustration, emergent magnetic monopoles and phase transitions[15-20], as well as for the implementation of multiple computation tasks such as Boolean logic operations[11-14,21] and neuromorphic computing[22-26]. Despite the different lateral coupling mechanisms available, including long-range dipolar coupling and nearest-neighbour chiral coupling, these are engineered either through the geometric design[27-30] or by locally tuning the magnetic anisotropy during the fabrication process[19]. As a consequence, the functionality of coupled nanomagnets is determined once fabricated and the subsequent electrical modulation at run-time remains challenging.

Here, we present a mechanism to realize electrically tunable lateral coupling between two adjacent Ising-like nanomagnets by exploiting the VCMA-mediated competition of the symmetric exchange interaction and Dzyaloshinskii-Moriya interaction (DMI) that favour the parallel (P) and antiparallel (AP) alignment of the nanomagnet magnetizations respectively (Fig. 1a). Employing this concept for extended networks, we are able to explore the voltage-controlled phase transition in artificial spin ices and to obtain an electrically programmable nanomagnetic Ising network that can serve as a neuromorphic computing element.

**Voltage-controlled magnetic coupling**

The voltage-controlled nanomagnets are fabricated from a Pt/Co-based multilayer that has a large interfacial DMI[31,32] and tunable perpendicular magnetic anisotropy[33]. As shown in Fig. 1b, the structure comprises two protected regions (in red) with a fixed out-of-plane (OOP) magnetic anisotropy and a 50 nm-wide gated region (in blue) with tunable magnetic anisotropy. In the gated region, the Co layer is exposed to the electrolyte $GdO_x$, constituting a solid-state ionic gate structure[34-36] and thus its magnetic anisotropy can be reversibly modulated between in-plane (IP) and OOP by applying positive and negative voltages to the top gate electrode, respectively. In contrast, in the protected region where an 8 nm-thick $SiO_2$ layer is inserted between the Co layer and $GdO_x$, the migration of oxygen ions from $GdO_x$ to the top interface of Co is blocked and hence its magnetic anisotropy is protected from voltage modulation. The detailed fabrication process is described in Methods and Supplementary Information S1.

The $GdO_x$ was grown in oxygen-deficient conditions such that, in the as-fabricated device, oxygen ions at the top interface of the Co layer in the gated region are partially absorbed by $GdO_x$. The gated region exhibits IP magnetic anisotropy whereas the surrounding protected regions have OOP magnetic anisotropy, giving an OOP-IP-OOP anisotropy configuration. Moreover, as a result of the interfacial DMI at the Pt/Co interface, the magnetization within the OOP-IP and IP-OOP transition regions twists following a left-handed chirality enforced via chiral coupling[19]. As a result, the two OOP magnetizations in the protected regions are effectively AP coupled (Fig. 1a). In order to obtain the AP ground state, an oscillating and decaying magnetic field is applied perpendicularly to the devices, serving as the demagnetization protocol (Supplementary Information S2). As shown in Fig. 1c (left panel), for an array of coupled elements (Fig. 1b), the magnetization in the two coupled OOP regions exhibits either ↑↓ or ↓↑ alignment. After applying a negative gate voltage $V_G$ of -2.5 V for 90 min, the gated region exhibits OOP anisotropy thanks to the formation of Co-O bonds promoting perpendicular magnetic anisotropy[33]. Due to the collinear alignment in the OOP-OOP-OOP configuration, the DMI influence is reduced and the symmetric exchange interaction results in ↑↑ and ↓↓ low energy states. In other words, the two OOP magnetizations in the protected regions are P coupled. The conversion from AP to P coupling is demonstrated with the demagnetized configurations given in Fig. 1c. This coupling can be described as an effective exchange energy

$$E = -J(\mathbf{S}_1 \cdot \mathbf{S}_2), \tag{1}$$

where $\mathbf{S}_1$ and $\mathbf{S}_2$ represent the direction of the adjacent Ising macrospins, which can only point ↑ or ↓, and $J$ is the coupling strength that can be tuned between P ($J > 0$) and AP ($J < 0$).

In order to systematically study the voltage-controlled coupling, nanomagnet elements were fabricated on a Hall bar for the electrical detection of the OOP magnetization via the anomalous Hall effect (Fig. 2a). Due to the large difference in size between the protected and the gate regions, the Hall resistance mainly reflects the state of the protected regions. Following the demagnetization protocol, we recorded hysteresis loops, which consistently show the switching between three magnetization levels, corresponding to ↑↑ (↓↓) at large positive (negative) fields and ↑↓ or ↓↑ at intermediate fields (Fig. 2b). Note that the Hall resistance starts at the middle resistance level corresponding to ↑↓ or ↓↑, indicating that the demagnetized magnetic configuration is AP. We then record minor loops to quantify the coupling strength between two protected regions. The minor loops are shifted horizontally by the exchange bias field $H_{\text{bias}}$ = 127±16 Oe (-130±16 Oe) when the loop starts from positive (negative) saturation fields, confirming the presence of the AP coupling (red loops in Fig. 2c). The AP coupling strength can be estimated from $J = H_{\text{bias}} M V_{\text{OOP}}$ = -2.5±0.3 eV, where $M$ and $V_{\text{OOP}}$ are the magnetization and volume of the magnetic material in the protected region, respectively. After applying a negative gate voltage, the magnitude of $H_{\text{bias}}$ gradually decreases and eventually its sign is inverted, indicating the conversion from AP to P coupling (orange and green loops in Fig. 2c). The magnitude of exchange bias with P coupling is similar to that of AP coupling, and the strength is estimated to be 2.2±0.3 eV. We then verified that the demagnetized magnetic configuration is P as the Hall resistance started from either the highest or the lowest resistance levels corresponding to ↑↑ or ↓↓ (Fig. 2c), confirming the AP-to-P coupling conversion. Additionally, the magnetic coupling can be reversibly converted between AP and P by changing the $V_G$ polarity. As shown in Fig. 2d, we altered the $V_G$ polarity back and forth 13 times and $H_{\text{bias}}$ changed signs accordingly. Note that the time for the first conversion (~90 min) is longer than the subsequent conversion times (~30 min), which is also reflected by the reversal of the blue and purple loops in Fig 2c. This could be related to the additional energy cost of the first detachment of the ions from their original positions as well as to the formation of ionic conduction paths that provide subsequent faster conversion[37].

When the coupling $J$ is close to zero, the magnetic configuration obtained after different demagnetization cycles exhibits a stochastic behaviour. As shown in Fig. 2e, we start from the P state where the percentage of obtaining AP alignment, $\mathcal{P}_{AP}$, is close to 0. When a gate voltage of $V_G = 2.5$ V is applied, $\mathcal{P}_{AP}$ gradually increases over time and approaches 1. The percentage of AP alignment follows the Boltzmann law and is given by:

$$\mathcal{P}_{AP} = \frac{e^{-J/k_B T_{eff}}}{e^{-J/k_B T_{eff}} + e^{J/k_B T_{eff}}}, \qquad (2)$$

where $k_B$ is the Boltzmann constant and $T_{eff}$ is the effective temperature resulting from the demagnetization protocol (Fig. S3). $\mathcal{P}_{AP}$ is then regulated by $V_G$ that modifies the coupling strength, giving $\mathcal{P}_{AP} > 0.5$ for $J < 0$ and $\mathcal{P}_{AP} < 0.5$ for $J > 0$. We note that $\mathcal{P}_{AP}$ remains almost unchanged after disconnecting the device for 8 hours, indicating that the coupling is non-volatile thanks to the ionic gating effect[8]. We then apply a $V_G$ alternating between ±3 V and observe a synaptic plasticity of $\mathcal{P}_{AP}$ that varies according to $V_G$ (Fig. 2e). We repeat the demagnetization process 130 times at $J < 0$, $J \approx 0$ and $J > 0$ (at I, II and III in Fig. 2e), and the magnetic configuration alternates between AP and P randomly with different percentages (Fig. 2f), which enables a modulation of the correlation between neighbour macrospins as required for Ising-type probabilistic computing[38-41].

**Mechanism for the AP/P coupling conversion**

To provide deeper insight into the mechanism of the AP/P coupling conversion, we consider a macrospin model and calculate the coupling strength in response to the magnetic anisotropy of the gated region. As shown in Fig. 3a, $S_1$ and $S_2$ represent the direction of the magnetization in the protected regions, while $S_g$ represents the direction of the magnetization in the gated region. The total energy of this system is given by the sum of the symmetric exchange energy ($E_{ex}$), antisymmetric exchange energy ($E_{DM}$) and the magnetic anisotropy energy ($E_{an}$) of the gated region, which can be written as:

$$E = E_{ex} + E_{DM} + E_{an} = -J_{ex}\sum_{<i,j>} \mathbf{S}_i \cdot \mathbf{S}_j - \mathbf{D}_{eff}\sum_{<i,j>} \mathbf{S}_i \times \mathbf{S}_j - K_g V_g S_{gz}^2, \quad (3)$$

where $J_{ex}$ and $\mathbf{D}_{eff}$ denote the effective exchange energy and the DMI vector. $J_{ex} > 0$ and $\mathbf{D}_{eff} < 0$ in Pt/Co with left-handed chirality. $<i,j>$ represents all the possible combinations of the nearest-neighbour pairs of macrospins. $V_g$ is the volume of the magnetic material in the gated region and $K_g$ is the effective anisotropy constant of the gated region, which is experimentally tuned between IP ($K_g < 0$) and OOP ($K_g > 0$) by applying a gate voltage. The energies for the AP and P configurations are then:

$$E_{AP} = \begin{cases} 2D_{eff}; & \text{when } K_g V_g < -D_{eff} \\ -\dfrac{D_{eff}^2}{K_g V_g} - K_g V_g; & \text{when } K_g V_g \geq -D_{eff} \end{cases} \quad (4)$$

and

$$E_P = \begin{cases} \dfrac{J_{ex}^2}{K_g V_g}; & \text{when } K_g V_g < -J_{ex} \\ -2J_{ex} - K_g V_g; & \text{when } K_g V_g \geq -J_{ex} \end{cases} \quad (5)$$

The $\mathbf{S}_1$-$\mathbf{S}_2$ coupling strength is then determined from the difference between the energies $E_{AP}$ and $E_P$ and is given by: $J = (E_{AP} - E_P)/2$. On increasing $K_g$, $J$ increases from negative to positive, reflecting the AP-to-P coupling conversion (Fig. S5). Notably, if the gated region is strongly IP ($K_g << 0$), $J \approx D_{eff} < 0$, whereas if the gated region is strongly OOP ($K_g >> 0$), $J \approx J_{ex} > 0$, so providing an intuitive picture for the AP/P coupling conversion resulting from the VCMA-mediated competition between the symmetric and antisymmetric exchange interaction.

In order to quantify the coupling strength, we carried out micromagnetic simulations using the MuMax3 code[42]. The initial magnetization in the protected regions is set to be ↑↓ (or ↑↑) with the cell magnetizations in the gated region pointing in random directions. The system is then allowed to relaxed until a stable magnetic state is reached. In the ↑↓ configuration, we find that a Néel domain wall separates the two protected regions, whose width decreases with increasing $K_g$. In the ↑↑ configuration, two 90° domain walls form, which eventually vanish as $K_g$ is increased (Fig. S7). The total energy as a function of $K_g$ is shown in Fig. 3b. The main features of the $E_{AP}$ ($E_P$)-$K_g$ and $J$-$K_g$ curves agree with those of the macrospin model (Fig. S5).

When $K_g < 0$, $E_{AP} < E_P$ and $J$ saturates at -3.0 eV. When $K_g > 0$, $E_{AP} > E_P$ and $J$ increases with $\sqrt{K_g}$ (Supplementary Information S3). The experimentally obtained coupling strength (-2.5 eV for AP coupling and 2.2 eV for P coupling) are also in good agreement with the values estimated from the micromagnetic model.

**Voltage-controlled phase transition in Ising artificial spin ices**

As the magnetization in the protected regions can only point ↑ or ↓, it is possible to create an array of nanomagnet elements that mimic artificial Ising systems and display magnetic phase transitions that occur on changing the coupling strength. We first construct a one-dimensional (1D) Ising chain of nanomagnets by repeating an alternating structure of protected and gated regions in a line (Fig. 4a). The AP coupling in the as-fabricated Ising chain leads to antiferromagnetic (AFM) order on demagnetization (Fig. 4b upper panel). After applying $V_G$ = -2.5 V to the chain for 90 min, all of the couplings are converted to P and the chain of nanomagnets exhibits ferromagnetic (FM) ordering on demagnetization (Fig. 4b lower panel). We also vary the width of the gated region $w_g$ and find that the degree of both AFM and FM orders gradually decreases with increasing $w_g$ (Fig. 4c). Here the degree of the magnetic order is evaluated by determining the nearest-neighbour correlation function:

$$<S_i S_{i+1}> = \frac{\sum_{<i,j>} S_i \cdot S_j}{N}, \qquad (6)$$

where $N$ represents the number of the nearest pairs of macrospins and the sum runs over all nearest-neighbour pairs. $<S_i S_{i+1}>$ is equal to 1 and -1 for perfect AFM and FM order in the 1D Ising chain, respectively.

We then construct a 2D Ising artificial spin ice (Fig. 4d). The demagnetized square lattice exhibits an AFM checkerboard pattern in the as-fabricated state (Fig. 4e)[19]. A phase transition from AFM to FM order can then be achieved by applying a negative gate voltage (Fig. 4e). This transition can again be quantified by calculating the nearest-neighbour correlation function $<S_i S_{i+1}>$. By altering the $V_G$ polarity, the magnetic order is switched between AFM and FM with $<S_i S_{i+1}>$ varying between positive and negative values (Fig. 4f and Fig. S9). The $<S_i S_{i+1}>$ for FM order is smaller than that for AFM order, implying that the P coupling is weaker than the AP coupling. This could be due to the fact that the dipolar interaction becomes considerable in extended lattices and inhibits the formation of the FM order (Supplementary Information S7).

**Programmable Ising network and its application to neuromorphic computing**

Taking advantage of the flexibility of voltage control, we can adjust the individual magnetic couplings in an Ising network independently, thus providing addressable control of the couplings and creating an electrically programmable array of coupled nanomagnets. To demonstrate this feature, we fabricated a four-spin chain element where every gated region has a dedicated electrode and can be controlled independently (Fig. 5a and 5b). Analogous to the encoding method used in a binary system, we define the AP coupling as binary "1" ($J < 0$) and P coupling as binary "0" ($J > 0$), whereas the positive and negative gate voltages are defined as binary "1" ($V_G = 2.5$ V) and "0" ($V_G = -2.5$ V), respectively. The four-spin chain element has $2^3 = 8$ coupling configurations ($J_1, J_2, J_3$) corresponding to "000", "001", "010", "011", "100", "101", "110" and "111", which can be programmed one-by-one through the three gate voltages ($V_1, V_2, V_3$) (Fig. 5c). For instance, if the gate voltages are all positive i.e., ($V_1, V_2, V_3$) = "111" in terms of the electric signal, all the couplings are AP i.e., ($J_1, J_2, J_3$) = "111" in terms of the coupling configuration. In this case, the demagnetized magnetic configuration has a high percentage of AP alignment for all nearest-neighbour pairs $S_1|S_2$, $S_2|S_3$ and $S_3|S_4$ (Fig. 5d). If $V_1$ is switched to negative then ($V_1, V_2, V_3$) = "011" and ($J_1, J_2, J_3$) = "011". Therefore, after demagnetization, the $S_1|S_2$ pair exhibits P alignments with a high percentage, while the $S_2|S_3$ and $S_3|S_4$ pairs remain AP aligned with a high percentage (Fig. 5e). In this manner, we can obtain all eight coupling configurations (Fig. S11).

By extending this scheme to a more complex 2D network, it is feasible to construct a programmable Ising network whose couplings can be electrically adjusted between AP and P. Such a network has applications for both Boolean and non-Boolean computing. In the Supplementary Information S11, we describe the realization of reconfigurable Boolean logic gates such as a controlled-NOT and a controlled-Majority gate (Fig. S14), in which a dual logic functionality can be implemented by controlling the polarity of the control voltage. Moreover, our approach offers an efficient way to build an Ising-type neural network whose vertices are magnetically coupled to each other and each coupling is electrically adjustable. Many combinatorial optimization problems that are ubiquitous in fields such as artificial intelligence, bioinformatics, drug discovery, cryptography, logistics and route planning, can be mapped to an Ising network with specific Hamiltonians[45,46]. The solution of such problems can be obtained by finding the spin alignment corresponding to the ground state of the Ising network. In spintronic-based neuromorphic computing schemes, however, the couplings between vertices in an Ising-type neural network are usually achieved with additional CMOS circuits or by resistive crossbar arrays[40,43,44]. To illustrate the capability of solving combinatorial optimization problems using a magnetically-coupled network, we experimentally solve a benchmarking Max-Cut problem in an eleven-spin Ising network (Fig. 6).

The Max-Cut problem is frequently used for circuit design and machine learning[47,48], and is one of the most basic combinatorial optimization problems. In a typical Max-Cut problem, one starts with a system (a graph) in which a certain number of elements (the vertices of a graph) are related to each other by pairwise connections (the edges of a graph) with assigned weights. Finding the solution to the Max-Cut problem consists of maximizing the total weight of the edges between two mutually exclusive subsets of vertices. In our implementation, the nanomagnets represent the vertices of the graph, which are separated into two sets according to their magnetization, ↑ or ↓. The coupling strengths $J_{ij}$ corresponds to the weight of the edges $w_{ij}$ in the graph. Solving the Max-Cut problem is equivalent to minimizing the energy of an Ising network with the same connections. Since we can electrically program the strength of each individual coupling, our nanomagnetic system can serve as a combinatorial optimization problem solver. In Figure 6a and 6b, a schematic of an eleven-spin Ising network is shown, which represents a specific Max-Cut problem where each nanomagnet has either two or three connections. As shown in Fig. 6c (upper panel), the demagnetized configuration of the Ising network whose couplings are all AP ($J_{ij} < 0$) reveals the solution for the Max-Cut problem of a graph whose edge weights are all positive ($w_{ij} > 0$). In order to change the weight $w_{34}$ to negative ($w_{34} < 0$), the corresponding coupling $J_{34}$ is tuned to P ($J_{34} > 0$) by applying $V_G$ = -2.5 V for 90 min and the demagnetized magnetic configuration gives the new solution (Fig. 6c, lower panel). Therefore, the solution of Max-Cut problems can be obtained by relaxing a physical system to its ground state within a finite time. Different Max-Cut geometries can be implemented using our approach (Figure S12). However, due to the geometric limitations of a 2D nanomagnetic network and the short-range nature of chiral coupling, only nearest-neighbour connections are possible. More general network structures can be implemented by exploiting the long-range dipolar interaction[23,49], or by electrically coupling distant elements using the spin-transfer torque effect[40] (Fig. S13). Moreover, the ability to electrically program the magnetic coupling permits the adjustment of the Max-Cut problem at run-time, so enabling hardware-level programmability of the solver.

**Conclusions**

We have shown that we can electrically tune the magnitude and sign of the lateral coupling between nanomagnets by taking advantage of the antisymmetric exchange interaction and modifying the magnetic anisotropy in a gate region between the nanomagnets. The change in magnetic anisotropy is a consequence of the electrochemical reaction localized at the interface and ion migration in the gate dielectric under an electric field. The time required for the modification is limited by the reaction rate and ionic mobility, and can be reduced, in principle, using a high-mobility ion conductor[50] or the electronic-version of the VCMA effect, which is capable of operating at GHz frequencies[7].

Our approach offers the possibility to investigate collective phenomena, such as the coupling-dependent phase diagram[51] and phase boundaries of mixed FM/AFM Ising-like artificial spin ices[52,53], as well as the exotic magnetic phase of a spin glass[18,20]. Moreover, we have provided proof-of-concept demonstrations of reprogrammable nanomagnetic Boolean logic gates and combinatorial Ising solvers, which will inspire future research on unconventional computing devices based on nanomagnets.

**Author Contributions:** Z.Luo, L.J.H. and P.G. conceived the work and designed the experiments; A.H. and Z.Liu fabricated the devices. C.Y. and Z.Liang performed the MFM and electric measurements with the support of Y.F.; C.Y. and Z.Liang analysed and interpreted the data with the help of M.H., Y.H. and J.Y.; C.Y. and L.W. performed the micromagnetic simulations; Z.Luo, P.G. and L.J.H. worked on the manuscript together. All authors contributed to the discussion of the results and the manuscript revision.

**Competing interests:** Authors declare no competing interests.

**Methods**

**Device fabrication**

Films of Ta (5 nm)/Pt (5 nm)/Co (1.5 nm)/Al (2 nm) were deposited on a 200-nm-thick $SiN_x$ layer on a silicon substrate using d.c. magnetron sputtering at a base pressure of <2.7×10$^{-6}$ Pa and an Ar pressure of 0.4 Pa during deposition. The Al layer was oxidized to induce perpendicular magnetic anisotropy in the Co layer using a low-power (30 W) oxygen plasma at an oxygen pressure of 1.3 Pa. The fabrication of the voltage-controlled coupled nanomagnets was carried out with electron-beam lithography. Continuous magnetic films were milled into the shape of the bottom electrodes with Ar ions through a negative resist (ma-N2401) mask. The upper Co/AlOx layers were milled through a positive resist [poly(methyl methacrylate), PMMA] mask to create the nanomagnets and lattice structures. Using electron-beam evaporation, a protective layer of Cr (2 nm)/SiO$_2$ (8 nm) was deposited, with the protected region defined using a lift-off process through a second PMMA mask patterned by electron-beam lithography. Then an electrolyte layer of GdO$_x$ (30 nm) was deposited using reactive magnetron sputtering at an Ar pressure of 0.4 Pa and with a mixed gas flow of 50 sccm Ar and 1 sccm O$_2$. In order to promote the ionic gating effect, a short milling process was implemented to partially remove the AlO$_x$ layer in the gated region. Finally, top electrodes of Cr (2 nm)/Au (3 nm) were fabricated using electron-beam lithography combining electron-beam evaporation with a lift-off process. The base pressure for the electron-beam evaporation was <1.3×10$^{-4}$ Pa and the deposition rate for Cr, SiO$_2$ and Au was 0.5 Å/s. The main steps of the device fabrication are shown in Fig.S1. The magnetic anisotropies in the protected and gated regions were confirmed with polar MOKE measurements (Fig. S1).

**MFM measurements**

The MFM measurements were performed with a Bruker Dimension Icon Scanning Station mounted on a vibration- and sound-isolation table using tips coated with CoCr. To minimize the influence of the stray field from the MFM tip during the measurements, low-moment MFM tips were adopted. We repeated the MFM measurements and found that the magnetization in the nanomagnets remains unchanged, confirming that the MFM tips do not alter the magnetic configurations. For the voltage-control of the coupled nanomagnets, they were mount on a dedicated holder and connected to a source meter (Keithley 2400) with wire bonding. The MFM images were captured after employing the demagnetization protocol (Supplementary Information S2). All of the MFM measurements were performed at room temperature and under ambient conditions.

**Electrical measurements**

For electrical measurements, the magnetic films were patterned onto a 1.5 μm-wide Hall cross using electron-beam lithography and the coupled nanomagnet elements were located in the centre of the Hall cross. The devices were then connected to a source meter (Keithley 2400) and voltmeter (Keithley 2182) with wire bonding. All of the electrical measurements were performed at room temperature and under ambient conditions.

**Micromagnetic simulations**

To understand the mechanism of the AP/P coupling conversion, micromagnetic simulations were carried out with the MuMax3 code[42] using a computation box containing $1{,}000 \times 1{,}000 \times 1$ cells with $2 \times 2 \times 1.5$ nm$^3$ discretization and the following magnetic parameters: saturation magnetization $M_S = 0.9$ MA m$^{-1}$, effective OOP anisotropy field in the protected region $H_{\text{eff}} = 500$ mT, exchange constant $A = 16$ pJ m$^{-1}$ and interfacial DMI constant $D = -1.5$ mJ m$^{-2}$.

**Figures and figure legends:**

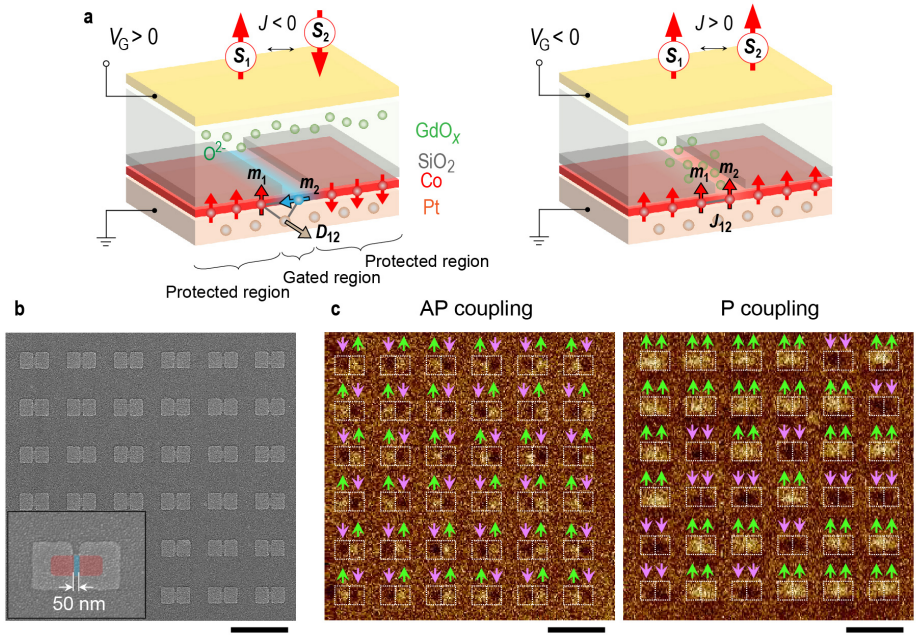

**Figure 1 | Basic concept of voltage-controlled magnetic coupling. a**, Schematics of coupled nanomagnet elements illustrating the principle of AP/P coupling conversion. On application of positive gate voltages, oxygen ions migrate away from the Co interface and the coupling between two protected regions is AP induced by interfacial DMI: $H_{DMI} = -\mathbf{D}_{12} \cdot (\mathbf{m}_1 \times \mathbf{m}_2)$, where $\mathbf{D}_{12}$ is the DM vector, and $\mathbf{m}_1$ and $\mathbf{m}_2$ are two nearest-neighbour magnetic moments. On application of negative gate voltages, oxygen ions migrate to the Co interface and the coupling becomes P due to the symmetric exchange interaction. **b**, SEM image of a 6×6 array of nanomagnetic structures. As shown in the inset, red and blue colours indicate the protected and gated regions respectively. **c**, MFM images of the 6×6 array with AP (left) and P (right) coupling. The bright and dark areas in the nanomagnet regions in the MFM images correspond to ↑ and ↓ magnetization, respectively. All the scale bars are 1 μm.

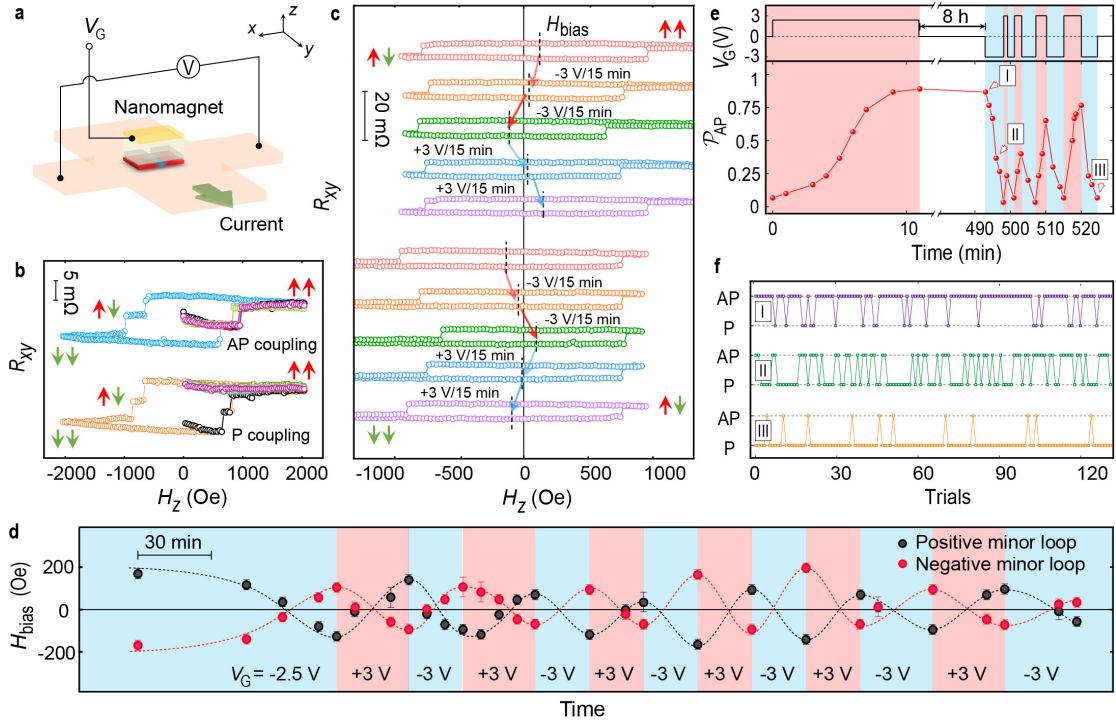

**Figure 2 | Reversible conversion of AP/P magnetic coupling. a**, Schematic of the Hall device used for electrical transport measurements. **b**, Magnetic hysteresis loops of a nanomagnet element on the Hall bar with AP (top) and P (bottom) coupling. The full hysteresis loops show three resistance levels corresponding to magnetic configurations of ↑↑, ↑↓ (↓↑) and ↓↓. The half hysteresis loops are measured after employing the demagnetization protocol and with the magnetic field starting from zero. **c**, Minor hysteresis loops with magnetic field starting from saturation. The magnitude of $H_{bias}$ is indicated with the black dashed lines. The $H_{bias}$ obtained from positive minor hysteresis loops (top 5 curves) is opposite to that obtained from negative minor hysteresis loops (bottom 5 curves). **d**, Evolution of $H_{bias}$ obtained from positive and negative minor hysteresis loops with respect to the gate voltage. The error bars represent the uncertainty in the estimation of $H_{bias}$. Red and black dashed lines are guides to the eye. **e**, Percentage of AP magnetic configurations obtained after demagnetization (lower panel) with the gate voltage given in the upper panel. Each percentage is determined from 30 trials carried out on one device. Red- and blue- shaded regions highlight the application of positive and negative gate voltages, respectively. **f**, Stochastic behaviour of the demagnetized configuration for different coupling strengths. The labels I, II and III correspond to the states indicated in **e**.

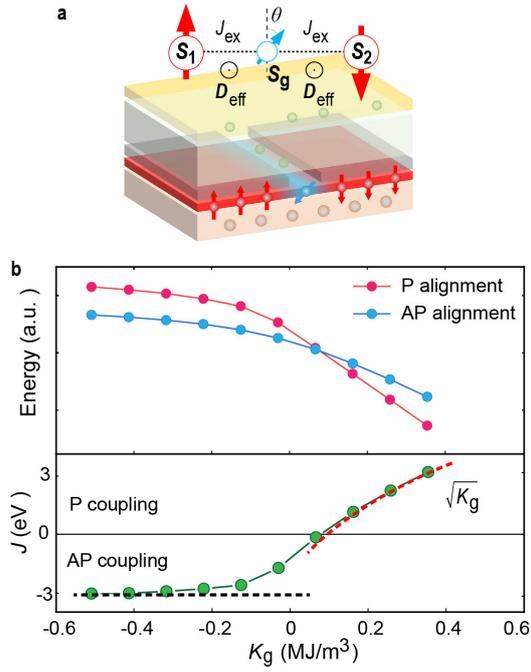

**Figure 3 | Macrospin model of AP/P coupling conversion and micromagnetic simulations of a nanomagnet element. a**, Schematic of macrospin model consisting of three macrospins representing the magnetizations in two protected regions and one gated region. **b**, Total energy for the AP (blue) and P (red) alignment as a function of $K_g$ determined from the micromagnetic simulations. The difference between the AP and P energies gives the coupling strength $J$ (green). The AP coupling strength saturates at -3.0 eV, as indicated by the horizontal black dashed line, whereas the strength of P coupling increases with $\sqrt{K_g}$, as indicated by the red dashed line.

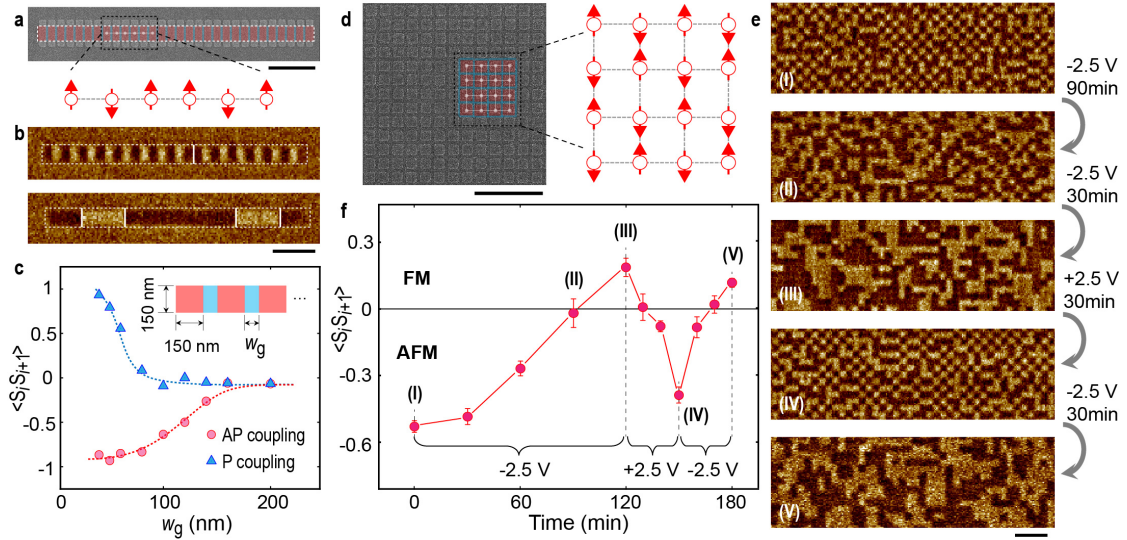

**Figure 4 | Voltage-controlled magnetic phase transition. a**, Coloured SEM image (top) and corresponding schematic (bottom) of a 1D Ising-like chain structure. The widths of the protected and gated regions are 150 nm and 50 nm. **b**, MFM images of the chain element with AP (top) and P (bottom) coupling. **c**, Nearest-neighbour correlation function $<S_iS_{i+1}>$ of AP and P coupling in a chain of 30 coupled regions as a function of gate width $w_g$. The dimension of the gated region is indicated in the inset. The red and blue dashed lines are guides to the eye. **d**, SEM image of the 2D Ising-like square lattice structure. Part of the image is indicated in colour (left) with a corresponding schematic (right). **e**, MFM image sequence of the 2D Ising-like square lattice showing reversible magnetic phase transitions between AFM and FM order. **f**, $<S_iS_{i+1}>$ as a function of gate voltage in the 2D square lattice. The states corresponding to the MFM image sequence (**I** to **V**) are indicated. The error bars represent the standard deviation of $<S_iS_{i+1}>$ evaluated from five 15×15 lattices. The bright and dark areas in the nanomagnet regions in the MFM images correspond to ↑ and ↓ magnetization, respectively. Red- and blue-shaded regions in the SEM images indicate the protected and gated regions. All the scale bars are 1 μm.

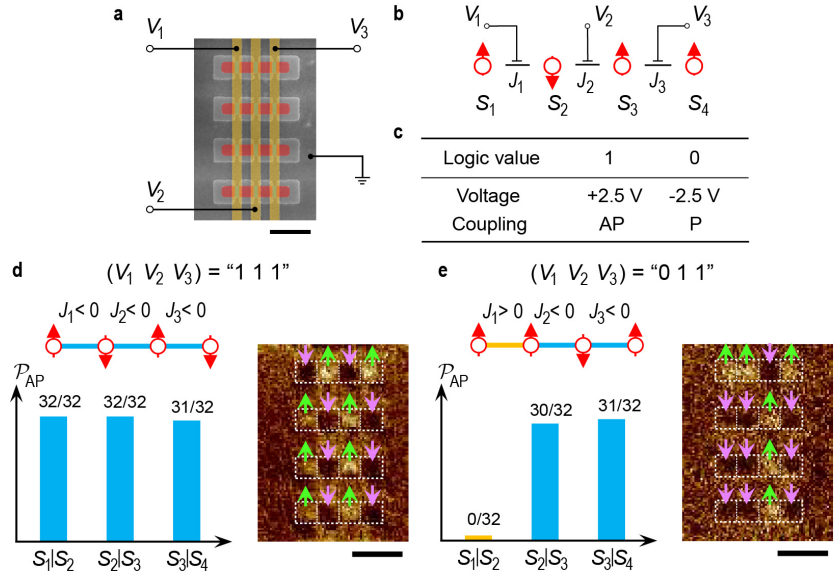

**Figure 5 | Addressable control of magnetic coupling in a four-spin chain. a** and **b**, Coloured SEM image and corresponding schematics of a programmable four-spin Ising chain. Red- and blue-shaded regions indicate the protected and gated regions, and yellow-shaded regions indicate the gate electrodes. **c**, Programming rules for the gate voltage and coupling. **d** and **e**, Coupling configurations for "111" (**d**) and "011" (**e**) programmed by applying the corresponding electric voltages. The blue and yellow connecting lines represent AP and P coupling, respectively. The percentages of AP alignment for the spin pairs $S_1|S_2$, $S_2|S_3$ and $S_3|S_4$ after demagnetization are shown (left), illustrating the programmed coupling configuration. Each percentage is obtained from the measurement of 32 elements. The MFM images of four element structures are shown with the bright and dark areas in the nanomagnet regions correspond to ↑ and ↓ magnetization, respectively, which is indicated with green and purple arrows (right). In order to guarantee the complete AP/P conversion, the gate voltages are applied for 90 min. All the scale bars are 500 nm.

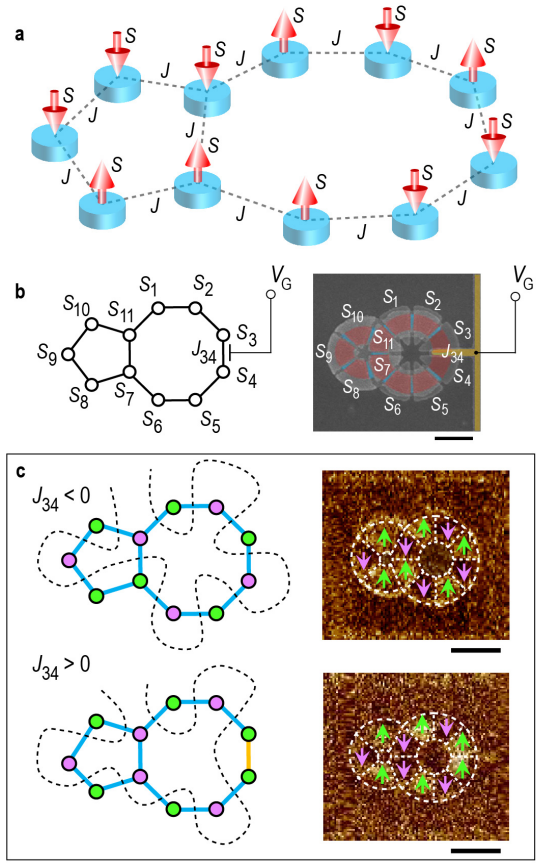

**Figure 6 | Programmable Ising network as hardware solver for Max-Cut problems.**
**a**, Schematic of a programmable Ising network based on coupled nanomagnets. Each coupling strength can be programmed by applying the corresponding electric voltage. **b**, Coloured SEM image and corresponding schematic of programmable eleven-vertex Ising network. In the SEM image, red- and blue-shaded regions indicate the protected and gated regions, while the yellow-shaded region indicates the gate electrode. **c**, Solutions to Max-Cut problem obtained from MFM images of demagnetized devices for the cases when $J_{34}$ is programmed to be AP (top) and P (bottom). The blue and yellow connecting lines in the schematics represent AP and P coupling. The black dashed line in each of the schematics indicates the cut lines separating vertices into two complementary sets (in green and purple), which is the solution to the Max-Cut problem with the corresponding weights. The bright and dark areas in the nanomagnet regions in the MFM images correspond to ↑ and ↓ magnetization, respectively, which is indicated with green and purple arrows. All the scale bars are 500 nm.

# Supplementary Information

# for

# Electrically programmable magnetic coupling in an Ising network exploiting solid-state ionic gating

Chao Yun[1,2,†], Zhongyu Liang[1,†], Aleš Hrabec[3,4,5], Zhentao Liu[3,4], Mantao Huang[6],

Leran Wang[1], Yifei Xiao[7], Yikun Fang[7], Wei Li[7], Wenyun Yang[1], Yanglong Hou[2], Jinbo Yang[1],

Laura J. Heyderman[3,4,*], Pietro Gambardella[5,*], Zhaochu Luo[1,*]

[1]State Key Laboratory of Artificial Microstructure and Mesoscopic Physics, School of Physics, Peking University, 100871 Beijing, China.
[2]School of Materials Science and Engineering, Peking University, 100871 Beijing, China.
[3]Laboratory for Mesoscopic Systems, Department of Materials, ETH Zurich, 8093 Zurich, Switzerland.
[4]Laboratory for Multiscale Materials Experiments, Paul Scherrer Institut, 5232 Villigen PSI, Switzerland.
[5]Laboratory for Magnetism and Interface Physics, Department of Materials, ETH Zurich, 8093 Zurich, Switzerland.
[6]Department of Materials Science and Engineering, Massachusetts Institute of Technology, Cambridge, MA, USA
[7]Division of functional Materials, Central Iron and Steel Research Institute Group, 100081 Beijing, China.
†These authors contributed equally: Chao Yun, Zhongyu Liang
*Correspondence to: zhaochu.luo@pku.edu.cn (Z.Luo); pietro.gambardella@mat.ethz.ch (P.G.); laura.heyderman@psi.ch (L.J.H.).

S1. Device fabrication and magnetic characterization

S2. Demagnetization protocol

S3. Details of macrospin and semi-micromagnetic model

S4. Micromagnetic simulations for different $K_g$

S5. Interplay between DMI and magnetic anisotropy

S6. Reliability of $<S_iS_{i+1}>$ obtained from different chips and positions

S7. Effect of dipolar interaction

S8. Programmable coupling configuration in a four-spin chain

S9. Programmable Ising networks for the 8- and 10-vertex Max-Cut problems

S10. Hybrid MTJ/Ising network structure for general Ising computing

S11. Reconfigurable nanomagnetic logic gates

# S1. Device fabrication and magnetic characterization

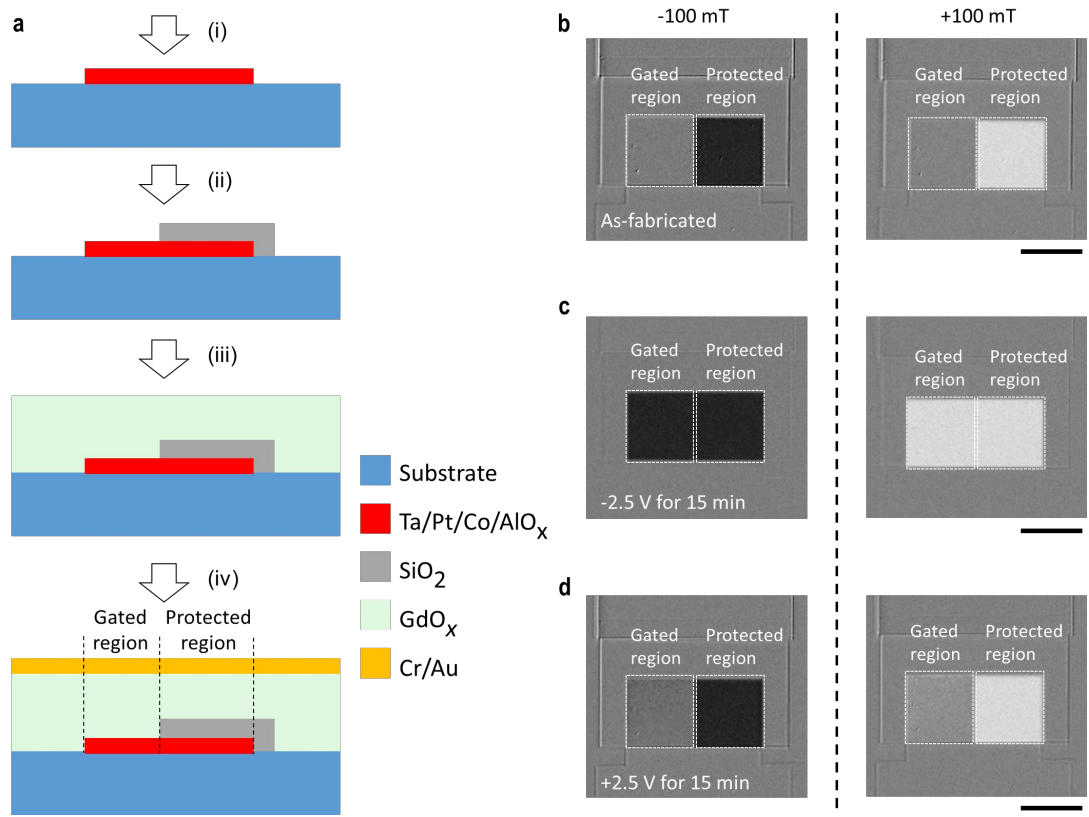

**Figure S1 | Device fabrication and magnetic characterization. a**, Schematic of main nanofabrication steps to create the electrically programmable coupled nanomagnets. (i) Ion milling of magnetic Ta (5 nm)/Pt (5 nm)/Co (1.5 nm)/AlO$_x$ (2 nm) multilayer, (ii) deposition and lift off to create a patterned protective layer of Cr (2 nm)/SiO$_2$ (8 nm), (iii) deposition of electrolyte layer of GdO$_x$ (30 nm) and (iv) deposition and lift-off to create a gate electrode of Cr (2 nm)/Au (3 nm). **b** to **d**, Polar MOKE images showing the evolution of OOP magnetic anisotropy in the gated and protected regions on application of a gate voltage: the as-fabricated state (**b**), after applying -2.5 V for 15 min (**c**) and after applying +2.5 V for 15 min (**d**). Each MOKE image is captured after saturating the sample with OOP magnetic fields of -100 mT (left) and 100 mT (right), and is subtracted from the image captured at saturation magnetic fields of 100 mT (left) and -100 mT (right). The white, grey and black contrast in the magnetic regions corresponds to ↑, IP and ↓ magnetization, respectively. All scale bars are 20 μm.

## S2. Demagnetization protocol

In order to obtain the low-energy magnetic configuration in an array of coupled nanomagnets, an oscillating magnetic field is applied perpendicular to the devices and the field amplitude is reduced over time (Fig. S2a). For this demagnetization protocol, the oscillating frequency of the magnetic field is 2 Hz (oscillation period $t_0 = 0.5$ s) and its amplitude is linearly reduced from $H_{max} = 200$ mT to zero with a constant step size of $\Delta H_{Demag} = 0.167$ mT. As shown in the MFM image of as-fabricated magnetic configuration of the Ising square-lattice element prior to applying magnetic fields (Fig. S2b), AFM-like domains are spontaneously formed, indicating the presence of AP coupling in the as-fabricated state. Following application of the demagnetization protocol, the size of the AFM-like domains increases, indicating that the array of coupled nanomagnets has been driven to a lower-energy state (Fig. S2c). Interestingly, on repeating the same demagnetization process, the AFM domain pattern changes, implying that the formation of AFM-like domains is stochastic.

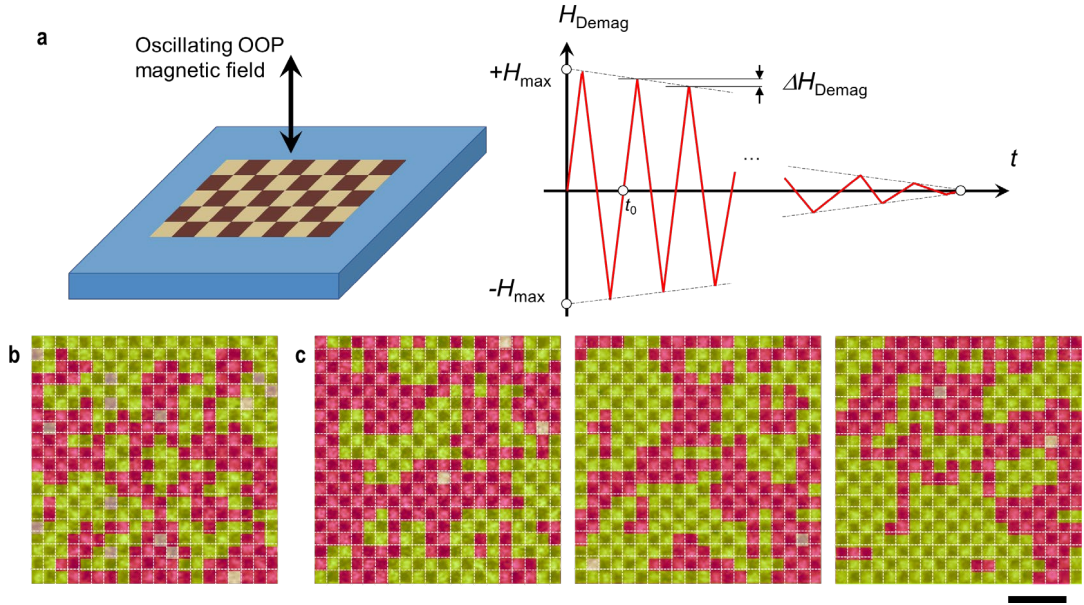

**Figure S2 | Demagnetization protocol. a**, Schematic showing the demagnetization procedure applied to the coupled nanomagnets. **b**, MFM image of the as-fabricated magnetic configuration of the Ising artificial spin ice prior to applying the demagnetization protocol. **c**, MFM images of the same area shown in **b** after applying the same demagnetization protocol. AFM-like domains are shaded in green and red. The bright and dark contrast in MFM images indicates nanomagnet regions with ↑ and ↓ magnetization, respectively. The scale bar is 1 μm.

To understand how the demagnetization protocol drives coupled nanomagnets to the low-energy state, we construct a square-lattice Ising macrospin model to simulate the demagnetization process. The macrospin approximation is used to model the thermally-active switching process with the switching probability $\mathcal{P}_{SW}$ given by[54]:

$$\mathcal{P}_{sw} = t_s f_0 e^{-\frac{E_b}{kT}}, \quad (S1)$$

where $t_s$, $f_0$, $E_b$, $k$ and $T$ represent the time to switch, the attempt frequency (~$10^9$ Hz), the energy barrier to switching, the Boltzmann constant and temperature, respectively. The switching energy barrier can be determined using the Stoner-Wohlfarth model where:

$$E_b = \begin{cases} E_{SW} + \dfrac{H_{eff}^2 m^2}{4E_{SW}} + H_{eff}m; & if\ |H_{eff}m| \leq 2E_{SW} \\ 2H_{eff}m; & if\ |H_{eff}m| > 2E_{SW} \end{cases}, \quad (S2)$$

when the effective magnetic field $H_{eff}$ is parallel to the direction of the original magnetization and

$$E_b = \begin{cases} E_{SW} + \dfrac{H_{eff}^2 m^2}{4E_{SW}} - H_{eff}m; & if\ |H_{eff}m| \leq 2E_{SW} \\ 0; & if\ |H_{eff}m| > 2E_{SW} \end{cases}, \quad (S3)$$

when the effective magnetic field $H_{eff}$ is antiparallel to the direction of the original magnetization. Here, the effective magnetic field $H_{eff}$ includes the external magnetic field $H_{ext}$ and the combined effect of the coupling with four nearest-neighbour sites:

$$H_{eff} = H_{ext} + \sum_{N.N.} J/m. \quad (S4)$$

The magnetic moment $m$ on the square-lattice site, the nearest-neighbour AP coupling strength $J$ (-2.5 eV) and the anisotropy-induced switching energy barrier $E_{sw}$ (15.7 eV) are all taken from the experimental results. Here, $J$ and $E_{sw}$ on each site is given by a Gaussian distribution to take into account the disorder in real devices. The magnetic configurations are obtained for different demagnetization step sizes $\Delta H_{Demag}$. The nearest-neighbour correlation $<S_iS_{i+1}>$ is determined to evaluate how close the magnetic configuration is to the ground state. The AP ground state on the square lattice is well-defined, forming a "checkerboard" pattern with $<S_iS_{i+1}>$ = -1. As shown in Fig. S3a, the magnetic configuration approaches the low-energy state on decreasing the demagnetization step size.

The demagnetization process can behave as a "thermal bath" that allows the array of coupled nanomagnets to relax into a low-energy configuration at an effective elevated temperature $T_{eff}$ [55,56]. We employ the Metropolis–Hastings algorithm to estimate $T_{eff}$ for our demagnetization protocol using the same coupling strength and distribution used in the macrospin model. The change of $<S_iS_{i+1}>$ with respect to the effective temperature parameter of $\beta J$ ($\beta = 1/kT$) is shown in Fig. S3b. A transition in $<S_iS_{i+1}>$ occurs around $\beta J \approx 0.5$, which agrees with the theoretical prediction of phase transition in the square-lattice Ising model:

$$(\beta J)_C = \frac{\ln(1+\sqrt{2})}{2} \approx 0.44. \tag{S5}$$

By comparing the values of $<S_iS_{i+1}>$ obtained using the macrospin model and using the Metropolis–Hastings algorithm, we can estimate the effectiveness of demagnetization protocol to drive the coupled nanomagnets to their ground state (Fig. S3c). The decrease of the demagnetization step size effectively decreases the temperature of "thermal bath" that saturates at a certain temperature ($\beta J \approx 0.44$) related to disorder in the coupled system. It also verifies the effectiveness of our demagnetization protocol with the experimental demagnetization step size of 0.167 mT being sufficient to realize the lowest-energy magnetic configuration.

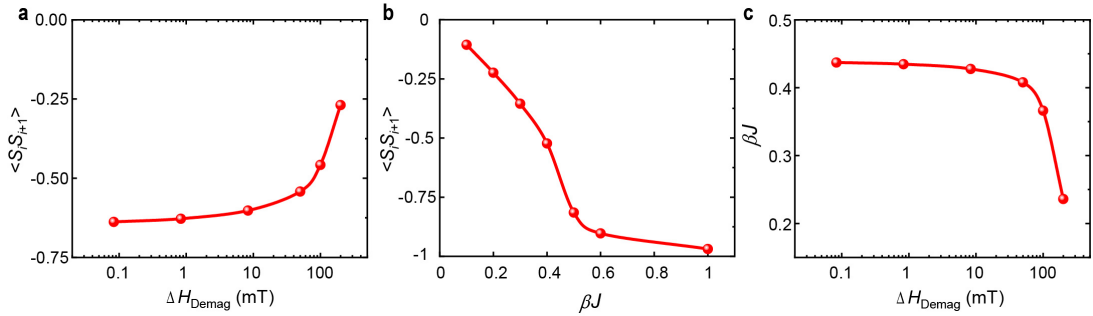

**Figure S3 | Simulation results of square-lattice Ising model. a**, $<S_iS_{i+1}>$ as a function of the demagnetization step size $\Delta H_{\text{Demag}}$. **b**, $<S_iS_{i+1}>$ as a function of the effective temperature $\beta J$ obtained with the Metropolis–Hastings algorithm for the square-lattice Ising model. **c**, Effective temperature parameter $\beta J$ as a function of the demagnetization step size $\Delta H_{\text{Demag}}$.

The protected regions are designed to have a high perpendicular magnetic anisotropy, which ensures that the magnetization is not perturbed by the stray field from the MFM magnetic tips during the measurements. This also means that the energy barrier for magnetization switching in the protected regions is higher than the thermal energy at room temperature. The coupled nanomagnet system is thus athermal and no thermally-active magnetization switching is observed during the experiments. Furthermore, the nearest-neighbour coupling strength is weaker than the energy barrier for switching the magnetization. Therefore, the voltage-controlled change of the coupling strength cannot induce the spontaneous switching of the magnetization without applying the demagnetization protocol.

As shown in Fig. S4a, the array of coupled nanomagnets exhibits an AFM-like pattern following demagnetization. The device was then exposed to a negative voltage of -2.5 V for 120 min, converting the coupling from AP to P. The same magnetic configuration was observed in the subsequent MFM measurement indicating that the energy barrier for switching the magnetization is higher than the coupling strength and the thermal energy (Fig. S4b). In order to demonstrate that the nearest-neighbour coupling has switched from AP to P, the array was demagnetized again and an FM-like pattern was observed, confirming the change of coupling from AP to P (Fig. S4c). Therefore, experimentally, demagnetization of the array is essential to show the conversion of the voltage-controlled coupling.

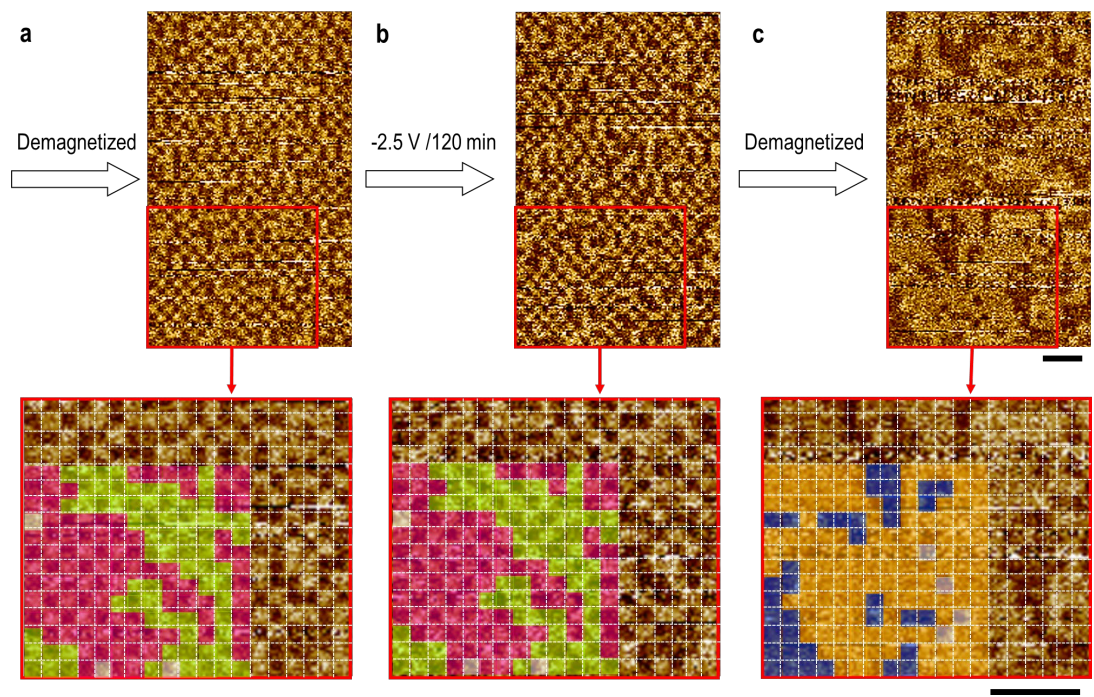

**Figure S4 | Demagnetized configurations following the electric gating. a**, MFM image of the magnetic configuration of the as-fabricated square-lattice array following demagnetization. **b**, MFM image of the magnetic configuration in the same area after applying a negative voltage of -2.5 V for 120 min. **c**, MFM image of the magnetic configuration in the same area following a second demagnetization. AFM-like domains are shaded in green and purple in the zoomed-in regions of **a** and **b**, and FM-like domains are shaded in yellow and blue in the zoomed in region of **c**. The bright and dark areas in the nanomagnet regions in the MFM images correspond to ↑ and ↓ magnetization, respectively. The scale bars are 1 μm.

## S3. Details of macrospin and semi-micromagnetic model

In this section, we will first give a more detailed description of the macrospin model that was briefly introduced in the main text and shown in Fig. 3a. We then turn to a semi-micromagnetic model to quantitatively estimate the coupling strength.

In the macrospin model, due to the strong OOP magnetic anisotropy, $S_1$ and $S_2$ can only point either ↑ or ↓. The tilt angle $\theta$ of $S_g$ is determined by minimizing the total energy. For AP alignment ($S_1 = \uparrow$ and $S_2 = \downarrow$), the energy can be written as:

$$E_{AP} = -J_{ex}\left[\cos\theta + \cos(\pi-\theta)\right] - D_{eff}\left[\sin\theta + \sin(\pi-\theta)\right] - K_g V_g \cos^2\theta \\ = -2D_{eff}\sin\theta - K_g V_g \cos^2\theta. \quad (S6)$$

When $K_g V_g < -D_{eff}$, $\sin\theta = -1$ i.e., $S_g = \leftarrow$ and $E_{AP} = 2D_{eff}$. When $K_g V_g \geq -D_{eff}$, $\sin\theta = D_{eff}/K_g V_g$ and $E_{AP} = -D_{eff}^2/K_g V_g - K_g V_g$. For P alignment ($S_1 = \uparrow$ and $S_2 = \uparrow$), and the energy can be written as:

$$E_P = -J_{ex}\left[\cos\theta + \cos\theta\right] - D_{eff}\left[\sin\theta + \sin(-\theta)\right] - K_g V_g \cos^2\theta \\ = -2J_{ex}\cos\theta - K_g V_g \cos^2\theta. \quad (S7)$$

Similarly, when $K_g V_g < -J_{ex}$, $\cos\theta = -J_{ex}/K_g V_g$ and $E_P = J_{ex}^2/K_g V_g$. When $K_g V_g \geq -J_{ex}$, $\cos\theta = 1$ i.e., $S_g = \uparrow$ and $E_P = -2J_{ex} - K_g V_g$.

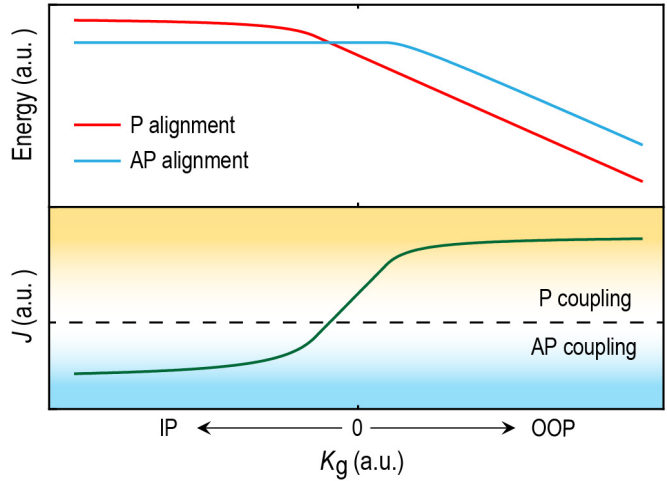

**Figure S5 | Energies for AP and P alignment, and coupling strength as a function of $K_g$ obtained from the macrospin model with $J_{ex}$ = 1.5 eV and $D_{eff}$ = -1 eV.**

In Pt/Co, the antisymmetric exchange interaction is weaker than the symmetric exchange interaction, i.e. $|J_{ex}| > |D_{eff}|$. The energy curves for the AP and P configurations are shown in Fig. S5. By determining the difference in energy between AP and P alignment, we can obtain the strength of the coupling between $S_1$ and $S_2$. As discussed in the main text, if the gated region

is strongly IP ($K_g \ll 0$), $J \approx D_{eff} < 0$, whereas if the gated region is strongly OOP ($K_g \gg 0$), $J \approx J_{ex} > 0$, so providing an intuitive picture for the AP/P coupling conversion resulting from the $K_g$-mediated competition between symmetric and antisymmetric exchange interaction.

Despite the fact that it is possible to explain the AP/P coupling conversion with the macrospin model, the "effective" interaction terms of $J_{ex}$ and $D_{eff}$ in Eq. 3 are not clearly related to the material parameters in real devices. To get closer to a real physical system, a semi-micromagnetic analytical model is developed on the basis of the macrospin model.

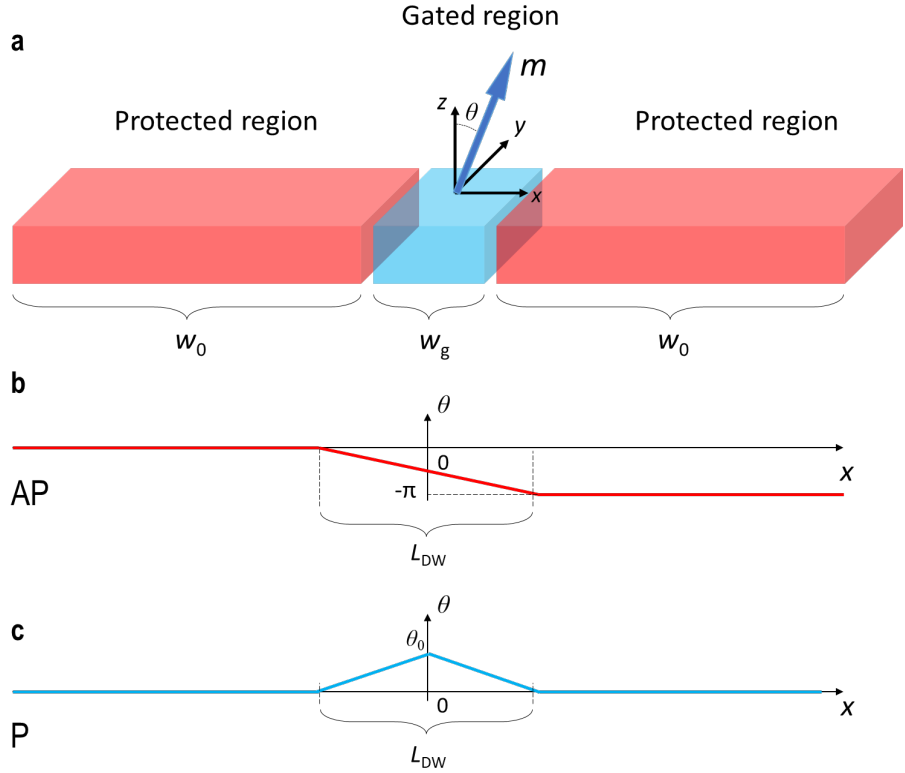

**Figure S6 | Semi-micromagnetic model. a,** Schematic of the basic element used for semi-micromagnetic model. **b** and **c**, Schematics of magnetization tilt angle $\theta$ for AP (**b**) and P (**c**) alignment of the magnetization in the neighbouring protected regions.

In the semi-micromagnetic model, the total energy including exchange energy, anisotropy energy and DMI energy, can be written as:

$$E = E_{ex} + E_{an} + E_{DM}$$
$$= \int dV A \sum_{i \in x,y,z} (\nabla \hat{m}_i)^2 + \int dV(-K\hat{m}_z^2) + \int -dVD\left[\hat{m}_z \nabla \cdot \hat{m} - (\hat{m} \cdot \nabla)\hat{m}_z\right], \quad (S8)$$

where $A$, $K$ and $D$ are the exchange energy constant, anisotropy constant and the DMI constant, respectively. $A$ and $D$ are constant throughout the magnetic regions, while $K$ is different in the protected and gated regions. We denote the anisotropy constant within the gated region as $K_g$ and as $K_0$ for the protected region.

For simplicity, we assume the magnetization lies in $xz$ plane and rotates linearly within domain walls. The structure of the basic element used for the model is shown schematically in Fig. S6a.

(i) For AP alignment ($S_1 = \uparrow$ and $S_2 = \downarrow$) (Fig. S6b), the boundary condition is $\theta|_{x=-w_0-\frac{w_g}{2}} = 0$ and $\theta|_{x=w_0+\frac{w_g}{2}} = -\pi$. Considering the continuous magnetization rotation and the symmetry of the structure, the magnetization at the centre can be either $\leftarrow$ or $\rightarrow$. However, due to the left-handed chirality in Pt/Co, the magnetization at the centre prefers to be $\leftarrow$ i.e., $\theta|_{x=0} = -\frac{\pi}{2}$. Hence the magnetization can be written as:

$$\hat{m} = [\sin\theta, 0, \cos\theta],$$

$$\theta = \begin{cases} 0; & \text{when } -\frac{w_g}{2} - w_0 \leq x \leq -\frac{L_{DW}}{2} \\ -\frac{\pi}{L_{DW}}x - \frac{\pi}{2}; & \text{when } -\frac{L_{DW}}{2} \leq x \leq \frac{L_{DW}}{2} \\ -\pi; & \text{when } \frac{L_{DW}}{2} \leq x \leq \frac{w_g}{2} + w_0 \end{cases}, \quad (S9)$$

where $L_{DW}$ represents the domain wall width.

Substituting Eq. S9 into Eq. S8, we obtain the following expression for the energy:

$$E_{AP} = \int A dx [(\frac{\partial \hat{m}_x}{\partial x})^2 + (\frac{\partial \hat{m}_z}{\partial x})^2] + \int -K dx \hat{m}_z^2 + \int -D dx \left( \hat{m}_z \frac{\partial \hat{m}_x}{\partial x} - \hat{m}_x \frac{\partial \hat{m}_z}{\partial x} \right)$$

$$= \int A \left(\frac{\partial \theta}{\partial x}\right)^2 dx + \int -K dx \cos^2\theta + \int -D dx \left(\frac{\partial \theta}{\partial x}\right) = E_{ex} + E_{an} + E_{DM}$$
(S10)

$$E_{ex} = \frac{AS\pi^2}{L_{DW}},$$

$$E_{an} = -2K_0 w_0 S - K_g w_g S + \frac{K_0 L_{DW} S}{2} + \frac{L_{DW} S \sin(2\pi w_g / L_{DW})}{2\pi}(K_g - K_0),$$

$$E_{DM} = \pi DS.$$

where $S$ is the cross-sectional area.

When $K_g < 0$, and after the total energy minimization with respect to $L_{DW}$, one finds: $L_{DW} = \pi\sqrt{\frac{2A}{K_0}}$. Since the gated region is narrow, we assume that $L_{DW} > w_g$ and $\sin(2\pi w_g / L_{DW}) \approx 2\pi w_g / L_{DW}$. Taking $L_{DW}$ into account, the energy expression becomes

$$E_{AP} = \pi S\sqrt{2AK_0} - 2K_0 w_0 S - K_g w_g S + S\sqrt{\frac{A}{2K_0}}\sin\left(w_g\sqrt{\frac{2K_0}{A}}\right)(K_g - K_0) + \pi DS. \quad \text{(S11)}$$

When the OOP magnetic anisotropy in the gated region is relatively strong ($K_g > 0$), the domain wall will be fully located in the gated region i.e., $L_{DW} < w_g$. The expression of energy can then be written as:

$$E_{AP} = \int AS\left(\frac{\partial\theta}{\partial x}\right)^2 dx + \int -KS dx \cos^2\theta + \int DS dx\left(\frac{\partial\theta}{\partial x}\right) = E_{ex} + E_{an} + E_{DM}, \quad \text{(S12)}$$

$$E_{ex} = \frac{AS\pi^2}{L_{DW}},$$

$$E_{an} = -2K_0 w_0 S - K_g w_g S + \frac{K_g L_{DW} S}{2},$$

$$E_{DM} = \pi DS.$$

By minimizing the total energy with respect to $L_{DW}$, we find: $L_{DW} = \pi\sqrt{\frac{2A}{K_g}}$. Then, introducing $L_{DW}$ into the energy density, we obtain:

$$E_{AP} = \pi S\sqrt{2AK_g} - 2K_0 w_0 S - K_g w_g S + \pi DS. \quad \text{(S13)}$$

(ii) For P alignment ($\mathbf{S}_1 = \uparrow$ and $\mathbf{S}_2 = \uparrow$) (Fig. S6c), the boundary condition is $\theta\big|_{x=-w_0-\frac{w_g}{2}} = 0$ and $\theta\big|_{x=w_0+\frac{w_g}{2}} = 0$. The magnetization can then be written as:

$$\hat{m} = [\sin\theta, 0, \cos\theta],$$

$$\theta = \begin{cases} 0; & \text{when } -\dfrac{w_{\text{g}}}{2} - w_0 \leq x \leq -\dfrac{L_{\text{DW}}}{2} \\ \dfrac{2\theta_0}{L_{\text{DW}}} x + \theta_0; & \text{when } -\dfrac{L_{\text{DW}}}{2} \leq x \leq 0 \\ -\dfrac{2\theta_0}{L_{\text{DW}}} x + \theta_0; & \text{when } 0 \leq x \leq \dfrac{L_{\text{DW}}}{2} \\ 0; & \text{when } \dfrac{L_{\text{DW}}}{2} \leq x \leq \dfrac{w_{\text{g}}}{2} + w_0 \end{cases} \qquad (S14)$$

According to the macrospin model, $\theta_0 = \pm\dfrac{\pi}{2}$ when $K_{\text{g}} \ll 0$ and $\theta_0 = 0$ when $K_{\text{g}} \gg 0$.

We first consider the case where $K_{\text{g}} \ll 0$ i.e., when the gated region is IP magnetized and $\theta_0 = \dfrac{\pi}{2}$. Substituting Eq. S14 into Eq. S8, we obtain the following expression for the energy density:

$$E_{\text{P}} = \int AS\left(\dfrac{\partial \theta}{\partial x}\right)^2 dx + \int -KS dx \cos^2\theta + \int DS dx \left(\dfrac{\partial \theta}{\partial x}\right) = E_{\text{ex}} + E_{\text{an}} + E_{\text{DM}}, \quad (S15)$$

$$E_{\text{ex}} = \dfrac{AS\pi^2}{L_{\text{DW}}},$$

$$E_{\text{an}} = -2K_0 w_0 S - K_{\text{g}} w_{\text{g}} S + \dfrac{K_0 L_{\text{DW}} S}{2} + \dfrac{L_{\text{DW}} S \sin(2\pi w_{\text{g}}/L_{\text{DW}})}{2\pi}(K_{\text{g}} - K_0),$$

$$E_{\text{DM}} = 0.$$

After minimizing the total energy with respect to $L_{\text{DW}}$, we find: $L_{\text{DW}} = \pi\sqrt{\dfrac{2A}{K_0}}$. Again, since the gated region is narrow, we assume $L_{\text{DW}} > w_{\text{g}}$ and $\sin(2\pi w_{\text{g}}/L_{\text{DW}}) \approx 2\pi w_{\text{g}}/L_{\text{DW}}$. Taking $L_{\text{DW}}$ into the energy density expression, we obtain:

$$E_{\text{P}} = \pi S\sqrt{2AK_0} - 2K_0 w_0 S - K_{\text{g}} w_{\text{g}} S + S\sqrt{\dfrac{A}{2K_0}} \sin\left(w_{\text{g}}\sqrt{\dfrac{2K_0}{A}}\right)(K_{\text{g}} - K_0). \quad (S16)$$

We then consider the case of $K_{\text{g}} \gg 0$ i.e., when the gated region is OOP magnetized and take $\theta_0 = 0$. Substituting Eq. S14 into Eq. S8, we obtain the following expression for the energy:

$$E_\text{P} = \int AS\left(\frac{\partial\theta}{\partial x}\right)^2 dx + \int -KSdx\cos^2\theta + \int DSdx\left(\frac{\partial\theta}{\partial x}\right) = E_\text{ex} + E_\text{an} + E_\text{DM}, \quad (S17)$$

$$E_\text{ex} = 0,$$

$$E_\text{an} = -2K_0 w_0 S - K_g w_g S,$$

$$E_\text{DM} = 0.$$

Hence,

$$E_\text{P} = -2K_0 w_0 S - K_g w_g S. \quad (S18)$$

By determining the energy difference between the AP and P configurations, we obtain:

$$E_\text{AP} - E_\text{P} = \pi DS < 0 \text{ when } K_g \ll 0 \quad (S19)$$

and

$$E_\text{AP} - E_\text{P} = \pi S\sqrt{2AK_g} + \pi DS > 0 \text{ when } K_g \gg 0. \quad (S20)$$

Comparing this with the results obtained from macrospin model, we obtain the relationship between physical parameters $D$ and $A$, and the "effective" interaction terms $D_\text{eff}$ and $J_\text{ex}$:

$$D_\text{eff} \approx \frac{\pi DS}{2} \quad (S21)$$

$$J_\text{ex} \approx \pi S\sqrt{\frac{AK_g}{2}} + \frac{\pi DS}{2}. \quad (S22)$$

The validity of these equations is confirmed by the good agreement of the coupling strength with that obtained from general micromagnetic simulations as described in main text. In particular, the magnitude of the AP coupling is given by: $J = D_\text{eff} \approx \frac{\pi DS}{2} = -3.3$ eV ($D = -1.5$ mJ/m$^2$ and $S$ = 150 nm × 1.5 nm), while the magnitude of P coupling is given by: $J = J_\text{ex} \approx \pi S\sqrt{\frac{AK_g}{2}} + \frac{\pi DS}{2} = 3.2$ eV ($A$ = 16 pJ m$^{-1}$, $K_g = M_SH_K/2$ = 0.9 MA m$^{-1}$ × 608.4 mT/2). In addition, we find that the P coupling strength has a $\sqrt{K_g}$ dependence (Fig. 3b). Since the value of $K_g$ tends to saturate at a certain value for $V_G < 0$, the coupling strength $J$ for the P coupling should have an upper limit.

## S4. Micromagnetic simulations for different $K_g$

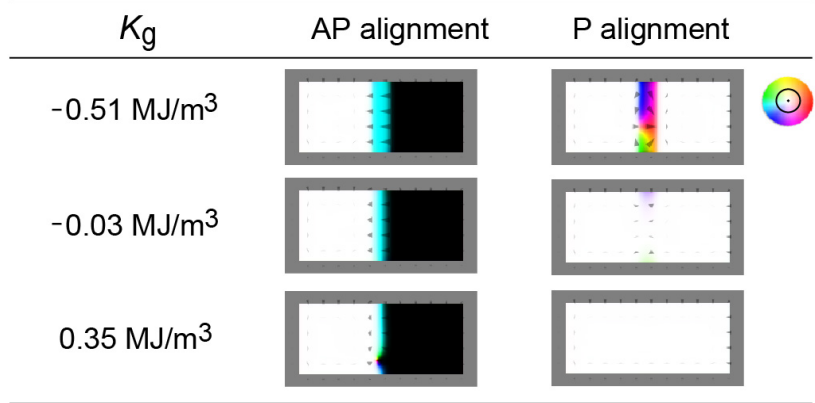

**Figure S7 | Snapshots of micromagnetic simulations for different $K_g$.** The direction of the magnetization is indicated by the colour wheel, and white and black correspond to ↑ and ↓ magnetization, respectively.

## S5. Interplay between DMI and magnetic anisotropy

In main text and Fig. 3, we present relationship between coupling strength and magnetic anisotropy in the gated region determined from micromagnetic simulations. The effective OOP magnetic anisotropy $K_{\text{eff}}$ used in the micromagnetic simulations is given by:

$$K_{\text{eff}} = K_u - \frac{\mu_0 M_s^2}{2}, \tag{S23}$$

where $K_u$ and $\mu_0$ are the interfacial uniaxial magnetic anisotropy constant and the magnetic permeability of free space, respectively.

In Fig. S8, we present further results from the micromagnetic simulations, highlighting the interplay between DMI and magnetic anisotropy in the gated region. When $K_g < 0$, the energy of the system for AP and P alignment is almost the same in the absence of DMI, which supports the fact that DMI is responsible for AP coupling. With increasing DMI, the difference in energy for AP and P alignment increases when $K_g < 0$, indicating the enhancement of AP coupling. This leads to an increase in the critical $K_g$ where the coupling is converted from AP to P. When $K_g > 0$, the energy of the system for AP alignment surpasses that for P alignment, resulting in a P coupling.

In the main text, we only consider the VCMA effect for the voltage control of the coupled nanomagnets. However, it has been reported that the DMI strength can be modified with electric fields and that the DMI decreases with decreasing OOP magnetic anisotropy since these two effects share the similar origin of spin-orbit coupling[57-60]. This may be partially the reason for the slight difference between experimental (-2.5 eV) and calculated values (-3.0 eV) of the AP coupling strength.

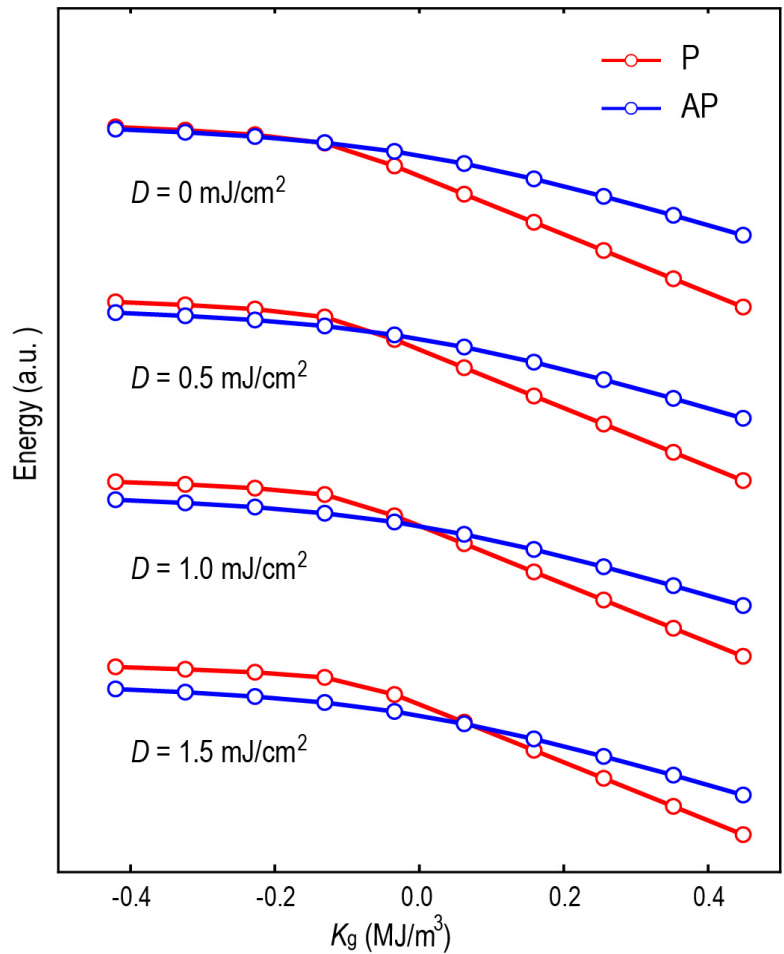

**Figure S8 | Energy of the system obtained from micromagnetic simulations for AP and P alignment as a function of $K_g$ for different DMI values.**

## S6. Reliability of <$S_iS_{i+1}$> obtained from different chips and positions

To illustrate the device-to-device reliability, the nearest-neighboring correlation function <$S_iS_{i+1}$> of the square lattice of 3 chips and 5 different devices per chip were measured as a function of the gate voltage (Fig. S9). The performance of the voltage-controlled magnetic coupling on changing the gate voltage is found to be robust with the trend in <$S_iS_{i+1}$> reproduced within <0.1 standard deviation on the same chip.

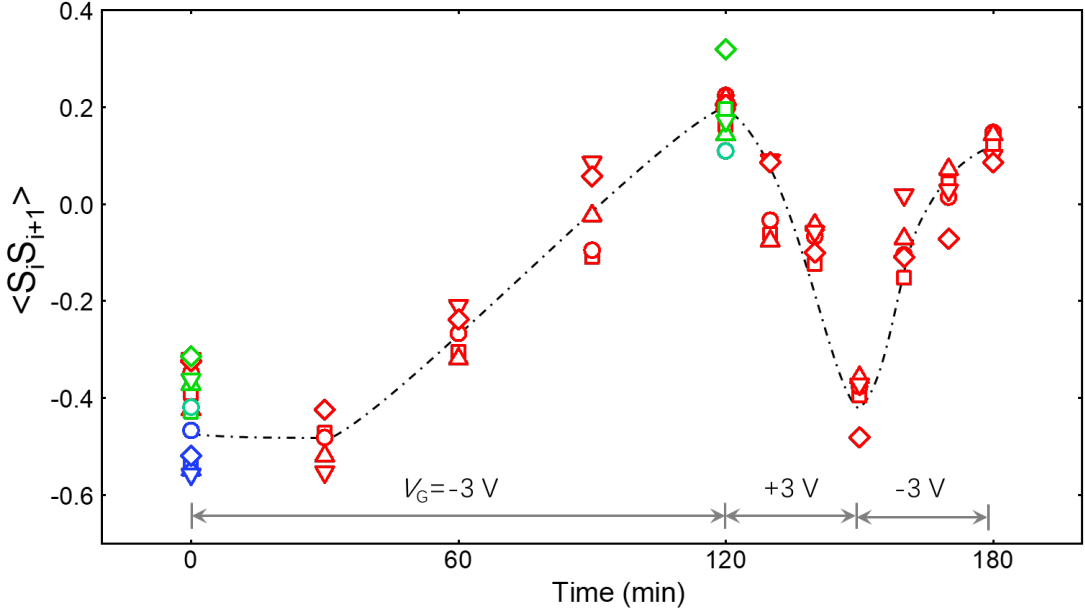

**Figure S9. <$S_iS_{i+1}$> as a function of gate voltage in 15×15 square lattices obtained from 5 devices fabricated on 3 different chips.** Red, green and blue colours indicate the results obtained from 3 different chips, while the different symbols indicate the results obtained from 5 different devices on the same chip.

## S7. Effect of the dipolar interaction

Here, we determine the effect of the dipolar interaction in the Ising artificial spin ice. First, we estimate the contribution of the dipolar interaction for the basic element with two protected regions (Fig. 1a and 1b). A rough estimation of the energy of the dipolar coupling between the two protected regions can be obtained by considering two point-like dipoles placed at the centre of each element at a distance $r$ from each other. In this case, the dipolar coupling $J_{dip}$ is given by:

$$J_{dip} = (E_{\uparrow\downarrow} - E_{\uparrow\uparrow})/2 = -\frac{\mu_0 m^2}{4\pi r^3} \approx -0.07 \text{ eV} \tag{S24}$$

with $m = 3.0\times10^{-17}$ A·m$^2$ (for nanomagnet dimensions of 150 nm × 150 nm × 1.5 nm) and $r = w_p + w_g = 200$ nm, where $w_p$ and $w_g$ are the widths of the protected and gated region, respectively. This dipolar coupling is almost two orders of magnitude smaller than the measured coupling of 2.5 eV as well as the estimated exchange-induced coupling of 3.0 eV. Therefore, the effect of the dipolar interaction is negligible in the basic element.

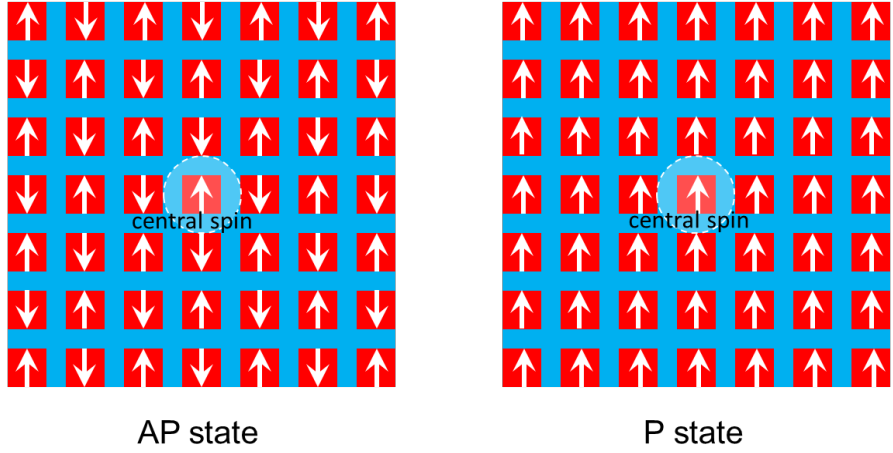

AP state        P state

**Figure S10 | Schematics of magnetic configurations in the AP and P state used to estimate the effect of the dipolar coupling in an Ising artificial square ice.**

We now estimate the dipolar interaction in the extended Ising artificial square ices. For this, we consider the dipolar and exchange interactions in square lattice shown in Fig. S10. As the dipolar interaction is a long-range interaction, the energy associated with it needs to take into account the interactions from the surrounding macrospins. We evaluate the effect of the dipolar interactions by calculating the energy difference when flipping the central macrospin $S_0$:

$$\Delta E = E\left(S_0 = \downarrow\right) - E\left(S_0 = \uparrow\right) = \Delta E_{dip} + \Delta E_{ex}. \tag{S25}$$

where $\Delta E_{dip}$ and $\Delta E_{ex}$ are the change in the dipolar and exchange energies on flipping the central spin. For simplicity, we consider the case where the surrounding macrospins are in the ground state. For AP coupling, the ground state has AFM order, and the change in energy due to the dipolar interaction on flipping the central macrospin is:

$$\Delta E^{AP}_{dip} = -\sum_i \frac{\mu_0 m^2 s_i}{2\pi r_i^3} \approx 0.38 \text{ eV} \tag{S26}$$

where $s_i$ and $r_i$ represent the orientation of the $i^{th}$ surrounding macrospins and the distance between the center and the $i^{th}$ macrospin, respectively. The dipolar energy is summed over all surrounding macrospins in a 100 × 100 square lattice similar to the experimental size. The dipolar interaction facilitates the formation of the AFM order. Due to the alternating up-down alignment of the magnetization for AP coupling, the energy change on flipping the central spin due to the dipolar interaction is small compared to $\Delta E_{ex} \approx 8J \approx 19.6$ eV. For P coupling, the ground state has FM order, and the energy change due to the dipolar interaction on flipping the central macrospin is:

$$\Delta E^{P}_{dip} = -\sum_{i} \frac{\mu_0 m^2 s_i}{2\pi r_i^3} \approx -1.29 \text{ eV} \tag{S27}$$

The dipolar interaction inhibits the formation of the FM order and, due to the uniform alignment of the macrospins in the P state, the energy difference induced by dipolar interaction becomes sizable.

In the experiment, we find that the strength of the AP and P coupling measured in the basic element is similar (Fig. 2d), while the correlation function $\langle S_i S_{i+1} \rangle$ for FM order is significantly smaller than that for AFM order in the extend structures (Fig. 4f). This could be due to the fact that the dipolar interaction becomes considerable in extended lattices and inhibits the formation of the FM order.

## S8. Programmable coupling configuration in a four-spin chain

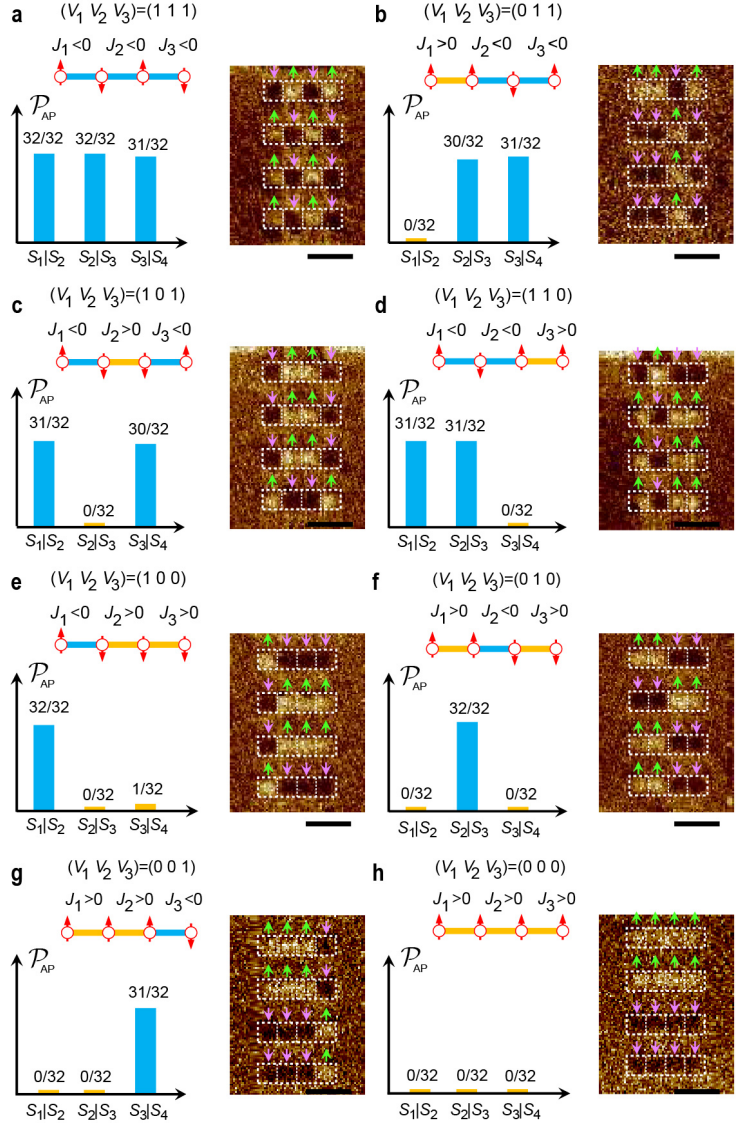

**Figure S11 | Programmable coupling configurations in a four-spin chain. a** to **h**, All $2^3 = 8$ coupling configurations that can be programmed using electric voltages. The applied voltages and corresponding coupling configurations, as well as one of the ground states, are shown. The blue and yellow connecting lines represent AP and P coupling, respectively. The probabilities of AP alignment for the pairs of $S_1|S_2$, $S_2|S_3$ and $S_3|S_4$ after demagnetization are shown (left), illustrating the programmed coupling configuration. Each percentage is obtained from the measurement of 32 elements. The MFM images of four selected element structures are shown with green and purple arrows indicating the magnetization of ↑ and ↓ respectively (right). In the MFM images, the bright and dark areas in the nanomagnet regions in MFM images correspond to ↑ and ↓ magnetization, respectively. In order to guarantee the AP/P conversion, the gate voltages are applied for 90 min. All the scale bars are 500 nm.

**S9. Programmable Ising networks for the 8- and 10-vertex Max-Cut problems**

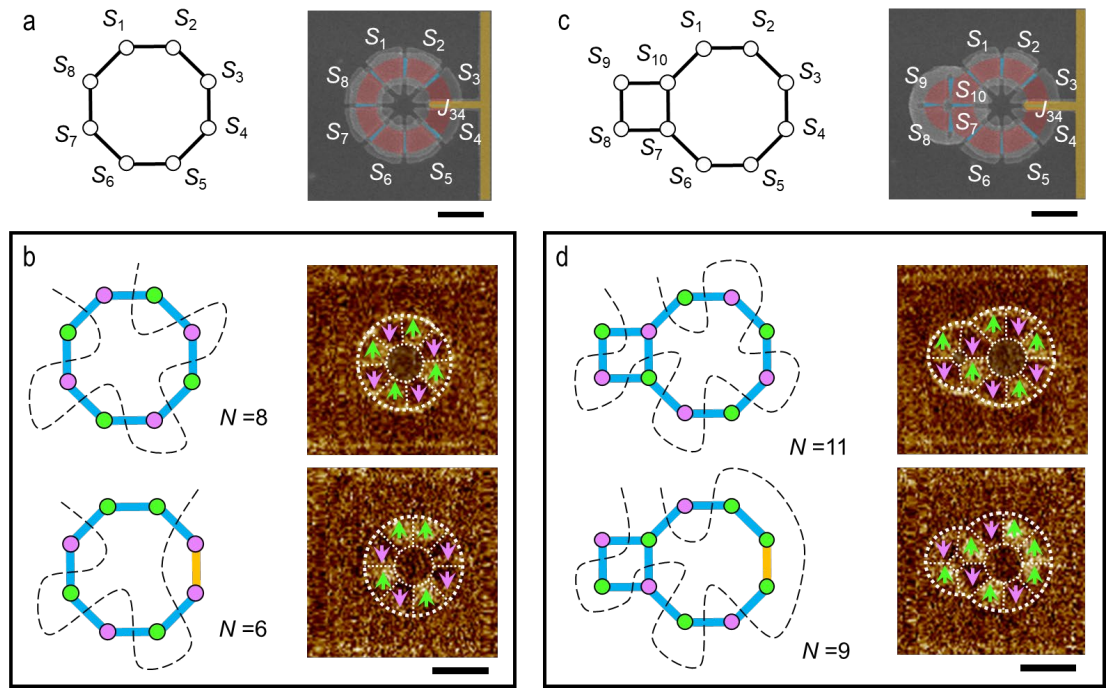

**Figure S12. Programmable Ising networks for the 8- and 10-vertex Max-Cut problems. a**, Schematic and coloured SEM image of a programmable 8-vertex Ising network. **b**, Solutions to Max-Cut problem obtained from MFM images of demagnetized devices for the cases when $J_{34}$ is programmed to be AP (top) and P (bottom). **c**, Schematic and coloured SEM image of programmable 10-vertex Ising network. **d**, Solutions to Max-Cut problem obtained from MFM images of demagnetized devices for the cases when $J_{34}$ is programmed to be AP (top) and P (bottom). The blue and yellow connecting lines in the schematics represent AP and P coupling. The black dashed line in each of the schematics indicates the cut lines separating vertices into two complementary sets (in green and purple), which is the solution to the Max-Cut problem with the corresponding weights. The bright and dark areas in the nanomagnet regions in the MFM images correspond to ↑ and ↓ magnetization, respectively, which is indicated with green and purple arrows. In the SEM image, red- and blue-shaded regions indicate the protected and gated regions, while the yellow-shaded region indicates the gate electrode. All the scale bars are 500 nm.

**S10. Hybrid MTJ/Ising network structure**

Due to the geometric limitations of the 2D Ising network, only nearest-neighbour connections are allowed, so giving a constraint on the complexity of the network. Here, we propose a hybrid MTJ/Ising network structure which can overcome this geometric constraint.

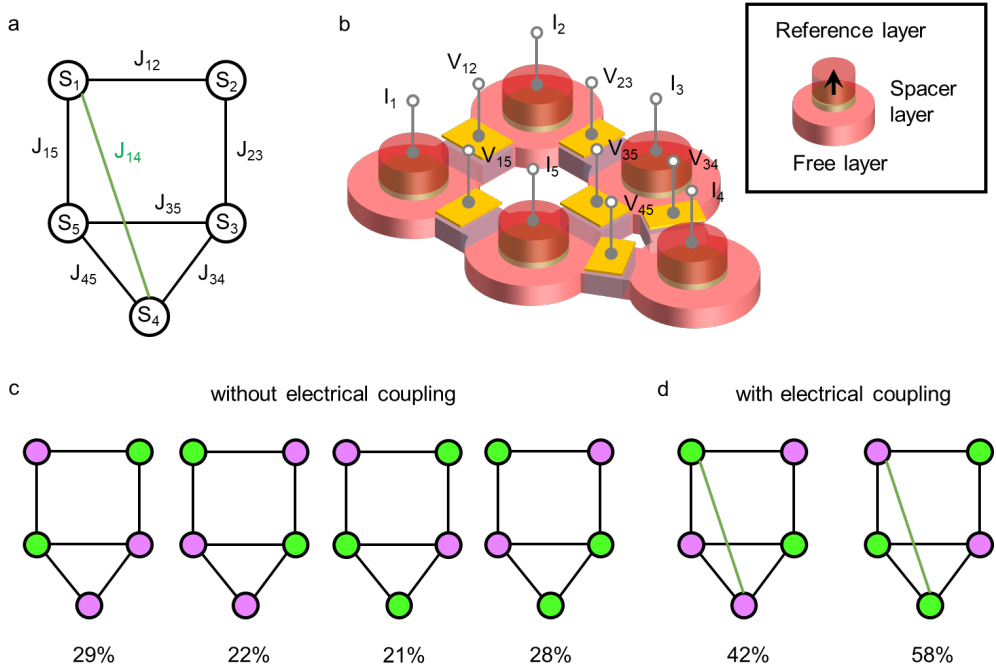

**Figure S13. Hybrid MTJ/Ising network structure for solving a complex Max-Cut problem. a,** Schematic of a 5-vertex Ising network. **b,** Schematic of the hybrid MTJ/Ising network structure. **c** and **d,** Simulation results of the ground state without (**c**) and with (**d**) electric coupling $J_{14}$. The percentages of the ground spin state obtained from 100 simulation trials are indicated.

To demonstrate a complex network with crossed connections, we present in Fig. S13a and S13b, an Ising-like nanomagnetic network including 5 vertices ($S_i$; $i$=1…5), 6 magnetic connections ($J_{12}$, $J_{23}$, $J_{34}$, $J_{45}$, $J_{35}$, $J_{15}$; $J_{ij}$ <0) and 1 electrical connection ($J_{14}$ <0). The nanomagnetic structure is similar to that shown in Fig. 6, which also contains two rings with an even and odd number of vertices. MTJs are fabricated on each spin vertex, and can be used to read and write the magnetization of the underlying Ising element (free layer) via the tunnel magnetoresistance effect and the spin transfer torque (STT) effect, respectively (Fig. S13b). Each MTJ is addressed by a current $I_i$. When $I_i$ is small, the STT effect is negligible and the magnetization of the Ising element (free layer) can be read via the tunnel magnetoresistance effect. When $I_i$ is large, the STT effect tends to switch the magnetization of the Ising layer parallel or antiparallel to the reference layer, depending on the polarity of $I_i$. Assuming the magnetization in the reference layer to be ↑, a positive (negative) $I_i$ gives an effective magnetic field pointing ↑ (↓) whose strength is determined by the intensity of $I_i$. In addition, the ionic gate structures are fabricated on each connection region (shown in yellow in Fig. S13b) and the gate voltage $V_{ij}$ is used to tune the magnetic coupling $J_{ij}$ between the vertices $S_i$ and $S_j$.

Coupling $J_{14}$ between vertices $S_1$ and $S_4$ is not possible in the 2D structure. Instead, we can couple vertex $S_1$ and vertex $S_4$ by applying electric currents through the MTJs on vertex $S_1$ and vertex $S_4$. The ground state of the Ising network is then obtained by applying the demagnetization protocol. The magnetization of $S_1$ and $S_4$ is read via the MTJ resistance with a small electric current. In addition, the electric currents

$$I_1 = I_{14}\text{sign}(S_4) \tag{S28}$$

and

$$I_4 = I_{14}\text{sign}(S_1), \tag{S29}$$

are injected to couple $S_1$ and $S_4$ via the STT effect. Here, $I_{14}$ is the intensity of the electric current corresponding to the coupling strength $J_{14}$ and sign($S_i$) determines the polarity of the electric current. As $J_{14}$ <0, $I_{14}$ <0 and the electric currents $I_1$ and $I_4$ cause the vertices of $S_1$ and $S_4$ to be AP. For example, when $S_1$ = ↑, $I_4=I_{14}$<0 and $S_4$ experiences an effective magnetic field pointing ↓, resulting in a current-induced AP coupling. This electrical process, including magnetization reading and electrical coupling, is continuously repeated throughout the demagnetization process to accomplish the integration of magnetic and electrical couplings.

In order to verify the effectiveness of the hybrid MTJ/Ising network structure, we performed a simulation with the same microspin model used for the square lattice. For the case of all magnetic couplings, the demagnetization process gives the four degenerate low energy spin states with approximately equal percentages as shown in Fig. S13c. In the presence of the electrical coupling $J_{14}$, the electric currents $I_1$ and $I_4$ effectively couple the magnetization direction of $S_1$ to that of $S_4$. In the simulations, we set the time period of updating the electric currents to be 1 ms, which is faster than the response time of magnetic field in our experiment. The STT-induced effective magnetic field is set to match the coupling strength of the magnetic coupling. After the demagnetization process, the simulation yields the doubly-degenerate low energy spin state with approximately equal percentages (Fig. S13d).

For a more general and complex Ising network, the vertex $S_i$ has $N_i$ virtual couplings interacting with $S^i_1…S^i_{Ni}$, and the electric current $I_i$ is given by:

$$I_i = \sum_{j=1}^{N_i} I_{ij}\text{sign}(S_j) \tag{S30}$$

where $I_{ij}$ is the intensity of the electric current corresponding to the coupling strength $J_{ij}$. During the demagnetization protocol, the magnetization of two spin vertices that are electrically coupled is read via the MTJ resistance with a small electric current. Then the electric currents are calculated and injected into the corresponding MTJ, leading to the "virtual" coupling.

As revealed above, in order to realize a complex Ising network with the functionality of programmability, the Ising network should contain (1) a gating structure for programmability and (2) an MTJ structure for the electric coupling. Both are compatible with state-of-art CMOS-back-end-of-line nanofabrication techniques. A similar hybrid structure of MTJs and coupled free layers has been demonstrated in previous experiments[62,63]. Further, the minimum feature size in our nanomagnetic device, *i.e.*, the width of the gated region, is 50 nm, which can be produced with large-scale nanofabrication of magnetic devices with good device reliability[61].

**S11. Reconfigurable nanomagnet logic gates**

Taking advantage of the binary nature of the Ising elements, a logic gate can be regarded as the ground state of a 2D Ising network with specific boundary conditions. When adjusting the logic inputs that determine the spin orientation of some specific vertices, the ground state configuration of the logic output vertices gives the result of the logic operation encoded in the network. As the logic operation only requires nearest-neighbouring interactions, our nanomagnetic structure provides a general physical platform to construct arbitrary logic circuits. By incorporating the voltage-controlled lateral coupling, the tunable magnetic coupling in our nanomagnetic structure allows for run-time reconfigurable logic operations. To demonstrate this capability, we have created a controlled-NOT gate, a fundamental Boolean logic gate (Fig. S14a). The dual operation functions of NOT and COPY can be interchanged according to the polarity of the gate voltage (Fig. S14b and S14c). When applying an electric voltage "1" to the gate electrode, the coupling is set to be AP and the direction of output magnetization is opposite to that of the input magnetization, thus accomplishing the NOT operation. If an electric voltage "0" is applied, the coupling is set to be P and the direction of the output magnetization is the same as that of the input magnetization, accomplishing the COPY operation.

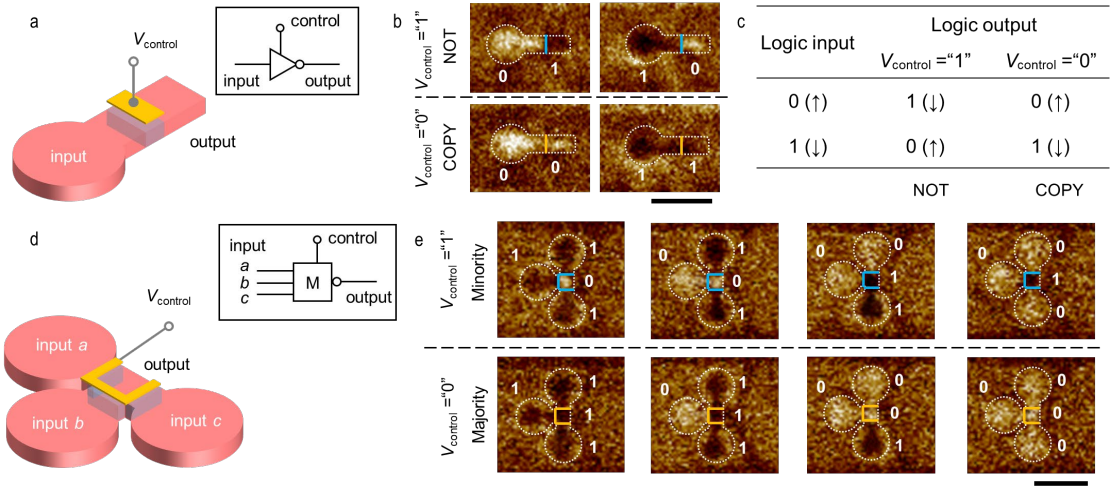

**Figure S14. Reconfigurable nanomagnetic logic gates. a**, Schematic of a controlled-NOT gate and corresponding logic circuit. **b**, MFM images of NOT operation for the AP coupling (top) and COPY operation for the P coupling (bottom). **c,** Truth table for the controlled-NOT gate. **d,** Schematic of a controlled-Majority gate and corresponding logic circuit. **e**, MFM images of Majority-NOT operation for the AP coupling (top) and Majority operation for the P coupling (bottom). In the schematics shown in **a** and **d**, red- and blue-shaded regions are the protected and gated regions, while yellow-shaded regions are the gate electrodes. The bright and dark areas in the nanomagnet regions in MFM images correspond to ↑ and ↓ magnetization, respectively. The blue and yellow lines in MFM images indicate the AP and P coupling. All scale bars are 500 nm.

Following the same principle, we can create a controlled-Majority gate, a functionally complete logic gate where any Boolean function can be implemented using a combination of Minority gates (Fig. S14d). The output of the controlled-Majority gate depends on the relative alignments of three inputs. As shown in Fig. S14e, if an electric voltage "1" ("0") is applied to the gate electrode, the coupling is set to be AP (P) and the direction of the output magnetization is opposite (equal) to that of the majority of the three input magnetizations, accomplishing the Minority (Majority) operation. Therefore, our logic scheme has the capability of dynamic reconfigurability, which increases the logic functionality of a device without increasing the number of logic gates and lead to a more compact logic chip.